%% file: ms.tex
\documentclass[draftclsnofoot,journal,onecolumn,12pt]{IEEEtran}
\pdfoutput=1 

\ifCLASSINFOpdf

\else

\fi
\usepackage{booktabs}
\usepackage{times,amsfonts}
\usepackage{graphicx}
\usepackage{epstopdf}
\usepackage[cmex10]{amsmath}
\allowdisplaybreaks[4]
\usepackage{array}
\usepackage{algorithm}
\usepackage{algpseudocode}
\usepackage[caption=false]{subfig}
\usepackage{color}
\usepackage{multirow}
\usepackage[table,xcdraw]{xcolor}
\usepackage{url,cite}
\usepackage{times}
\usepackage{bm}
\usepackage{amsmath}
\usepackage{amssymb}
\usepackage{amsfonts}
\usepackage{diagbox}
\usepackage{tabularx}
\usepackage{subfloat}
\usepackage[normalem]{ulem}
\usepackage{pifont}% http://ctan.org/pkg/pifont
\newcommand{\cmark}{\ding{51}}%
\usepackage{algorithmicx}
\usepackage{algpseudocode}

\algdef{SE}[DOWHILE]{Do}{doWhile}{\algorithmicdo}[1]{\algorithmicwhile\ #1}%

\onecolumn
\begin{document}

\title{Optimal Energy Management Strategies in Wireless Data and Energy Cooperative Communications}
\author{\IEEEauthorblockN{Jun~Zhou, 
		 Xiaodai~Dong,~\IEEEmembership{Senior~Member,~IEEE} \\ and Wu-Sheng~Lu,~\IEEEmembership{Fellow,~IEEE}, \\
        Department of Electrical and Computer Engineering\\
       University of Victoria, Victoria, BC, Canada\\
       E-mail: junzhou@uvic.ca, xdong@ece.uvic.ca, wslu@ece.uvic.ca}}

\maketitle

\begin{abstract}
This paper presents a new cooperative wireless communication network strategy that incorporates energy cooperation and data cooperation. The model establishment, design goal formulations, and algorithms for throughput maximization of the proposed protocol are presented and illustrated using a three-node network with two energy harvesting (EH) user nodes and a destination node. Transmission models are established from the performance analysis for a total of four scenarios. Based on the models, we seek to find optimal energy management strategies by jointly optimizing time allocation for each user, power allocations over these time intervals, and data throughputs at user nodes so as to maximize the sum-throughput or, alternatively, the minimum throughput of the two users in all scenarios. An accelerated Newton barrier algorithm and an alternative algorithm based on local quadratic approximation of the transmission models are developed to solve the aforementioned optimization problems. Numerical experiments under practical settings provide supportive observations to our performance analysis. 
\end{abstract}


\input{Introduction}

\input{Sys_Model}
\input{maxthrpt}
\input{algorithm2}
\input{simulation}
\input{Conclusion}

\begin{small}
\bibliographystyle{myIEEEtran}
\bibliography{Ref}
\end{small}

\end{document}

%% file: Introduction.tex
\section{Introduction}\label{sec:introduction}
%While the concept of power transmission by radio waves is not new \cite{history}, it is only recently, with rapid development and deployment of various wireless networks, contemporary radio wave energy harvesting (EH) techniques have emerged as promising solutions in the dealings with crucial energy constraint in battery-run networks. 
Energy harvesting (EH) has attracted significant interests from academia and industry in recent years, as both the concept of green communication and practical needs from low power sensor networks are growing steadily. There are successful industrial EH products that harvest energy through light, kinetic motion, pressure, etc., for sensors and controls \cite{industry}. Harvesting natural energy, e.g., solar, piezoelectric, kinetic and wind energy \cite{EHwsn}, in general presents a reliability challenge because of the instability of the natural environment. There is much work in the literature on natural energy harvesting. In \cite{policies}, two energy management strategies are proposed -- one is to maximize the total throughput within a given time slot, and the other is to minimize the transmission completion time given target data throughputs. In \cite{course516}, optimal online and offline energy management policies are designed by applying dynamic programming and staircase water-filling algorithm, respectively. Relative to natural resources of which EH can be greatly affected by environment changes, RF signals are more stable and human-controllable \cite{rfsurvey}. With rapid development and deployment of various wireless networks, contemporary radio wave energy harvesting techniques have emerged as promising solutions in the dealings with crucial energy constraint in battery-run networks. There are mainly two categories of RF wave based energy supply techniques. The first category is communication systems with simultaneous wireless information and power transfer (SWIPT) first proposed in \cite{swipt}. Because information and energy receivers are sensitive to different power values, two co-located receiver schemes for SWIPT, known as time switching (TS) and power splitting (PS), are described in \cite{swipt}. Based on TS and PS receiver schemes, a general PS scheme termed dynamic PS (DPS) is proposed in \cite{DPS,DPS2}. The second category is the RF-energy based wireless powered communication networking (WPCN) where wireless devices are charged by RF signals from a hybrid access point (H-AP) with power supply, first proposed in \cite{wpcn}. A ``harvest-then-transmit'' protocol is introduced in \cite{wpcn}, where the sum-throughput maximization and common-throughput maximization are investigated. Recently, WPCN has emerged as an appealing candidate solution in future self-sustainable wireless communication systems \cite{wpcnoverview,wpcnchallenge}. 

%The authors of \cite{weiheng} extend EC to a two-node scenario where the EH transmitter and EH receiver can transfer a part of the stored energy to each other.
Furthermore, efforts have been made to design cooperative and efficient energy management strategies based on a three-node network including a relay, where both the EH source and EH relay harvest energy from nature in \cite{energycooperation,cooperative,relay_DC,twoway_relay}. In \cite{energycooperation}, an energy cooperation strategy is introduced that allows the EH source node to share some energy with the EH relay node. A wireless cooperative transmission scheme with \emph{energy salvage} is proposed in \cite{cooperative}, where the source can harvest energy from the relay signal transmitted to the destination. Under the decode-and-forward (DF) relaying scheme in \cite{relay_DC}, two optimal power allocation algorithms are developed to solve the throughput maximization problems with delay-constrained and no-delay-constrained traffic at the destination, respectively. Furthermore, short-term sum-rate maximization problems are solved in \cite{twoway_relay}, where a two-way half-duplex relay channel is considered with DF relaying scheme.

Based on SWIPT techniques, the TS and PS protocols are also extended to three-node communication networks with relay assistance in \cite{swiptrelay,AFDFrelay,DF_multi}. Considering amplify-and-forwarding (AF) relaying in \cite{swiptrelay}, the EH relay node has no other energy sources but uses TS or PS to split the received RF signal from the source into two streams, one for EH and the other for information forwarding. Considering both AF and DF schemes, the performance of EH and throughput are analyzed in \cite{AFDFrelay}, where the relay node uses TS to split received RF signals. Furthermore, under the DF relaying strategy, multiple source-destination pairs with only one relay are considered in \cite{DF_multi}, where the EH relay adopts PS to harvest energy from the RF signals transmitted from multiple sources. On the other hand, relay assistance is also applied to improve the performance of WPCN. In \cite{htc}, ``harvest-then-transmit'' is extended to  ``harvest-then-cooperate'' scheme that enables a relay node to forward the information transmission of the source node. Different from \cite{htc} where the relay node only transmits the information of the source, in \cite{usercooperation} the ``relay'' node (the near user to H-AP) transmits its own data and forwards the information of the ``far'' user, which is called ``user cooperation''. 

%In a wireless sensor network for Internet of things, sensor nodes are constrained in their power supply and therefore harvesting energy from the environment is a natural choice. 
Energy harvesting becomes even more relevant in a wireless sensor network for the rapidly increasing Internet of Things (IoT) applications. Many of these low power sensor nodes are distributed in areas with no power supply and expected to last for years running on battery and/or on EH from all available sources.
Because nodes are located in different places, the energy they can harvest may differ as well. For example, a node that resides at a windy spot will harvest more wind energy than a neighboring node surrounded by structures. Therefore, not only data cooperation such as relaying will help to improve the overall performance of the network, RF energy cooperation/sharing will be an important approach to enhance the network operation. Moreover, due to the broadcast nature of wireless, the RF energy from one transmitting node can be harvested by all nodes nearby, regardless if the information is intended for the nodes. It is intuitive that the optimal energy and data cooperation strategy depends on the EH conditions of the nodes and the propagation channels among the nodes. Another consideration is that the order in which the individual nodes transmit their messages with continuous EH supplies appear to impact the system's performance, because the nodes who transmit later may store more energy, with the potential to perform better than the scenario when they transmit earlier. In this paper, we are motivated to investigate the energy and data cooperation strategies in a multi-node sensor network, where each node harvests energy from natural environments as well as from each other. 
%propose a new cooperative relay assisted wireless communication network protocol by harvesting energy from natural resources as well as ambient RF energy sharing, which exploits advantages of both data and energy cooperation. 
To enable tractable initial study, we consider a simple three node topology to illustrate the main idea, with potential extensions to more sophisticated network architectures in the future. The main contributions of the paper are summarized below.
%While energy management strategies based either on natural energy or on RF energy sources have been studied extensively in the literature, not much work has been devoted to strategies for harvesting energy from both natural and RF energy sources. For instance, RF signals transmitted by H-AP are considered as the only energy source for EH receivers in \cite{wpcn,htc,usercooperation,swipt,swipt22,DPS,DPS2,relay_DC,swiptrelay,DF_multi,AFDF_relay,twoway_relay}. However, applications exist when wireless user terminals want to remain operational when they are far away from H-AP. 
\begin{itemize}
	\item A new cooperative wireless communication network strategy is proposed. The model establishment, design goal formulations, and algorithms for throughput maximization under the proposal are carried out and illustrated using a three-node network with two EH user nodes and a destination node. Transmission models are established based on performance analysis for a total of four scenarios, each with two cases to distinguish which of the two users transmits first, that take into account node activity status in terms of whether it transmits (we then call it active) while harvesting surrounding natural energy, or it does not transmit (then call it inactive) while harvesting natural energy as well as RF signals broadcast by the active user. In addition, the power-splitting strategy adopted by the protocol allows a near-to-destination user to properly split the power of the RF signals from the ``far'' user, with one part for EH and the rest for relaying message to the destination. Collectively, this is a suite of transmission models for wireless networks with EH capabilities, where optimal throughputs can be achieved with the appropriate network cooperation strategy adaptive to the energy harvesting and propagation environments.
	%network cooperation and adaptability are enhanced to maintain satisfactory throughput in highly dynamic environment.
	\item Based on the proposed models, the problems of jointly optimizing the distribution of time intervals for data cooperation and energy harvesting, power allocations over these time intervals, and data throughputs at user nodes so as to maximize the sum-throughput or, alternatively, the minimum throughput of the two users are formulated as convex constrained problems in all scenarios. 
	\item Two fast algorithms are proposed to solve the aforementioned optimization problems. To be specific, an accelerated Newton barrier (NB) algorithm is developed to solve the constrained optimization problem, and the acceleration is achieved by a tailored line search technique which is required in each NB iteration. An alternative and often more efficient algorithm is developed based on a local approximation of the logarithmic terms involved. The results are convex quadratically constrained quadratic programming (QCQP) problems for several complex scenarios and quadratic programming (QP) problems for the rest of scenarios. Primal-dual path-following interior-point algorithms with closed-form line steps are deduced for these QCQP and QP problems and the efficiency of their implementations is evaluated and compared favorably with several available computer codes. 
\end{itemize}

The rest of the paper is organized as follows. Section \ref{sec:sys_model} presents the system model that consists of channel model, EH model, and transmission models for four possible scenarios. Section \ref{sec:maxthrpt} presents problem formulations based on the four scenarios, with the objective of maximizing the weighed sum-throughput and minimum throughput of the two users, respectively. In Section \ref{sec:algorithm}, we describe two convex optimization algorithms to solve the problems formulated in Section \ref{sec:maxthrpt}. Numerical results are given in Section \ref{sec:simulation} to evaluate how a wireless network in question performs with respect to the variations of the system parameters. Observations made from the simulation results are found supportive to our analysis. Finally, Section \ref{sec:conclusion} concludes the paper. We conclude this section with a remark on notation: In what follows, $\log(\cdot)$ denotes the base-2 logarithm, and $\mathbb{E}[\cdot]$ denotes the statistical expectation.
%A multi-access relay model is considered in \cite{mac}, the source nodes and relay nodes can transfer energy to one another in order for the overall sum-rate of the network.

%Taking into account wireless powered communication network (WPCN), \cite{2user-hap} proposes user cooperation protocol in the WPCN where the user nearer to the hybrid access point (H-AP) with a better channel helps relay the far user’s information to the H-AP, in order to achieve balanced throughputs.
%\cite{theoryswipt} has characterized fundamental tradeoffs between EH and information transfer. 
%

%However, those users who have harvested more energy with relatively small data amount to transmit indeed have surplus energy, while other users may be lack of energy. Furthermore, energy can be more efficiently utilized if the same information are transmitted through better channel conditions. To address these problems so that balance the communication qualities between multi-users and efficiently utilize the harvested energy, we need to apply adaptive strategies to power allocation and information transmission. 

%% file: Sys_Model.tex
\section{System Model}\label{sec:sys_model}
%We consider an EH network system consisting of two EH sensor users $U_1,U_2$ and a destination (D), . Note that the destination has a stable energy supply while two users are powered by the surrounding environment and ambient RF signals. The destination and two users are equipped with a single antenna, separately. When the active user is transmitting its information, the other inactive user harvests energy from the RF signals broadcast by the active user and from the surrounding environment. 
%If the user is nearer to the destination, it will separate the information signals and energy signals from the ``far'' user's transmission by PS as the decode-and-forward (DF) relay node, named ``data cooperation'' (DC). We assume that all channel conditions and energy arrival conditions are known before transmission.
\begin{figure}[t]
	\centering
	\includegraphics[width = 0.6\textwidth]{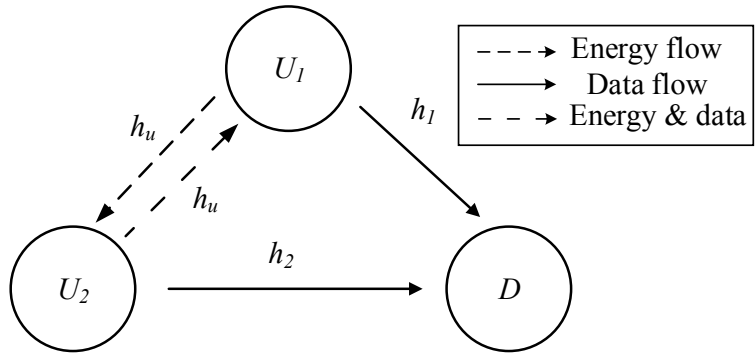}
	\caption{System model for a cooperative network with data cooperation and energy cooperation.}
	\label{fig:model}
\end{figure}

As illustrated in Fig. \ref{fig:model}, we consider a three-node network with two EH sensor nodes (called users) $U_1,U_2$ and a destination node $D$, each of which is equipped with a single antenna. Let $d_i$ $(i=1,2)$ be the distance from $U_i$ to $D$. Without loss of generality, we assume $d_1<d_2$, and name $U_1$ and $U_2$ the \textit{near user} and \textit{far user}, respectively. Concerning the power supply status in the network, node $D$, simulating a gateway or data collector in a wireless sensor network, is powered by a stable energy supply, while the two sensor nodes are powered by natural energy resources (e.g., solar, wind) as well as RF signals. The near user $U_1$ has data to send to the destination D, while the far user $U_2$ sends its data either directly to the destination or with the help of $U_1$ acting as a relay. The near user $U_1$ in our model acts as a decode-and-forward (DF) relay node and this message relaying process is called \emph{data cooperation} (DC). Without data cooperation, $U_2$ must send with higher transmission powert in order to reach $D$, compared with the relay mode. Data cooperation by $U_1$, however, will consume energy from $U_1$ and it is not directly obvious which method will result in higher sum throughput for $U_1$ and $U_2$ and lower total energy consumption. In this paper, both users take advantage of RF energy harvesting whenever the other user is transmitting. There are two cases in this scenario. When the data information from one user is intended for the other user, the concept of SWIPT may be applied as the same transmitted signal can be utilized for both data and power transfer. When the transmitted information signal from one user is not intended for the other user, the RF signal can still be harvested for energy by the other user due to the broadcasting nature of wireless channels. This process can be referred to as energy salvaging. Both cases of harvesting energy from RF signals broadcast by other users are referred to as \emph{energy cooperation} (EC). In the SWIPT case, when the near user $U_1$ receives a signal from the far user $U_2$, it may split the signal into two parts according to a PS strategy, with one part for EH and the other part for relaying the message from the far user $U_2$ to $D$. 
\subsection{Channel Model}
Referring to the system model in Fig. \ref{fig:model}, there are four channels in the network. Each channel can be characterized by a complex random variable $\tilde{h}$ with channel power gain $h=|\tilde{h}|^2$, which takes into account path loss and effects due to shadowing and channel fading. For simplicity, here we only consider the distance-dependent path loss so that the channel power gain is modeled as $h=\lambda d^{-\alpha}$, where $d$ is the distance from the transmitter to the receiver, $\alpha$ a path-loss exponent and $\lambda$ an average signal power attenuation at a reference of 1 unit of distance.

We use $h_i=|\tilde{h}_{i}|^2$ $(i=1,2)$ to denote the power gain of the uplink channel from user $U_i$ to $D$, and use $h_{ij}=|\tilde{h}_{ij}|^2$ $(i,j=1,2,i\neq j)$ for the power gain of the channel from $U_i$ to $U_j$. Throughout we assume time-division multiple access (TDMA) is used for the data transmission among the nodes, hence the channel reciprocity holds, i.e., $h_{12}=h_{21}$, which we shall denote by $h_u$ for simplicity. 

All channels in our system model are quasi-static block-fading, thus the channel gains remain constant during each transmission block $T$, but may vary from one block to another. For convenience, $T=1$ unit of time is assumed without loss of generality.
\subsection{Energy Harvesting Model}
\begin{figure}[t]
	\centering
	\includegraphics[width = 0.6\textwidth]{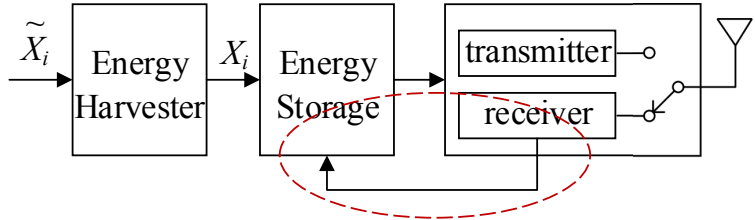}
	\caption{EH in a user node: $\tilde{X}_i$ is the rate of ambient natural energy arriving at user $U_i$ and $X_i$ the corresponding charging rate. The flow in the red circle indicates that the energy harvested from the received RF signals is also stored in the storage device before use, details of this are illustrated in Fig. \ref{fig:nodes_circuit}.}
	\label{fig:circuit}
\end{figure}

Fig. \ref{fig:circuit} illustrates the energy harvesting and energy flow in a user node. The charging rate for user $U_i$ is modeled as $X_i=\beta \tilde{X}_i$ with $0\leq \beta \leq 1$ for $i=1,2$, where $\tilde{X}_i$ denotes the rate of natural energy arriving at user $U_i$ and $\beta$ is a constant specifying charging efficiency. In what follows, we assume that 1) the charging rates are equal to the associated arriving energy rates, i.e., $\beta=1$; 2) the charging rates $X_i$ remain constant during each transmission block $T$; 3) ideal storage devices are used to store the harvested energy without leakage so that the harvested energy is used fully for data transmission; and 4) both user nodes are able to transmit or receive signals and harvest ambient natural energy simultaneously. 

\begin{figure}[h]
	\centering
	\includegraphics[width = 0.55\textwidth]{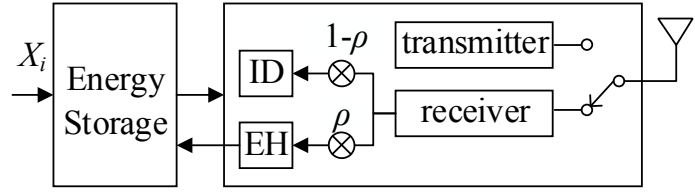}
	\caption{Receiver in a user node harvests and stores energy from RF signals.}
	\label{fig:nodes_circuit}
\end{figure}

Following the setup for SWIPT techniques \cite{swipt22}, the RF signals transmitted by the user nodes are used for information and energy transfer. As shown in Fig. \ref{fig:nodes_circuit}, the received RF signal with power $P$, is split into two parts: $\rho P$ with $0\leq \rho\leq 1$ for EH and the rest $(1-\rho)P$ for information decoding (ID). 
Two extreme circumstances are $\rho=1$ for the EC-only case with all received RF signals being used for EH, and $\rho=0$ for the DC-only case with all received RF signals being used for ID. An example of a possible scenario is that in a time duration $t$ ($0 < t < T$) within a transmission block of length $T$, an inactive user harvests $E_n$ from the natural energy and $E_h$ from RF signals with average transmission power $P$ through a channel with power gain $h$. In such a case, we have
\begin{subequations}
\begin{align}
E_n&=Xt\\ \label{EH_equ}
E_h&=\eta \rho P h t
\end{align}
\end{subequations}
where $\eta$ is the energy harvesting efficiency \cite{wpcn}.
\subsection{Transmission Models}
In order to be adaptive in different environmental circumstances, we propose to choose an optimal strategy from four scenarios: 1) Both DC and EC are applied;
%, as described in the beginning of this section (Fig. \ref{fig:model}); 
2) Only DC is applied - the near user $U_1$ relays the information from $U_2$ to $D$, but no energy is harvested from RF signals; 3) Only EC is applied - all received RF signals from the other user are harvested but no data relay; 4) Neither DC nor EC occurs - each user is powered by natural energy and transmits information to $D$ directly. Moreover, in each scenario we consider two cases: Case A when $U_1$ transmits first and Case B when $U_2$ transmits first. The node that transmits later will have more energy harvested for use. In the next section, by formulating and solving optimal power and time allocation for the eight combinations of energy-data cooperation and user node transmission order, the final optimal strategy will be obtained. In this section, the data transmission and energy harvesting protocol for each scenario and case are detailed. In the rest of the paper, we use S1-A to represent case A of Scenario 1, etc.

Similar to the harvest-then-transmit protocol \cite{wpcn}, for all four scenarios, we assume a possible initial time interval of length $t_0$ with $0<t_0<T$ when both users harvest natural energy before transmission. 

\textit{Scenario 1}: Both DC and EC are applied, see Fig. \ref{fig:model}. 

\textit{Case A}: $U_1$ transmits first.

% Please add the following required packages to your document preamble:
% \usepackage{multirow}
% \usepackage[table,xcdraw]{xcolor}
% If you use beamer only pass "xcolor=table" option, i.e. \documentclass[xcolor=table]{beamer}

Each transmission block $T$ consists of four non-overlapping time intervals of lengths $\{t_i,i=0,1,2,3\}$. In the rest of this paper, we use $P_a^{(b)}$ to denote the average transmission power by sensor node $U_b$ with $b = 1,2$ during $t_a$ with $a = 1,2,3$. The ID and EH activities at receivers in S1-A during time intervals $t_1,t_2$ and $t_3$ are summarized in Table \ref{P1-CA}. As a simple explanation of the table, during $t_1$, $U_2$ harvests the RF energy transmitted by $U_1$, hence in the $U_1 \to U_2$ column the box EH is marked with a \cmark. Since $U_2$ is inactive, the boxes in the $U_2 \to D$ column and $U_2 \to U_1$ are in gray color with a backslash $\backslash$.

During $t_{1}$, $U_1$ broadcasts RF signals to $D$ and $U_2$ with average power $P_1^{(1)}$, as well as harvests energy from natural resources. The complex baseband signal transmitted by $U_1$ is modeled as $x_1$ with $\mathbb{E}[|x_1|^2]=P_1^{(1)}$, which is received by $D$ and $U_2$ as $y_{10}$ and $y_{12}$, given by
\begin{subequations}
	\begin{align}
		y_{10}=\sqrt{h_{1}}x_{1}+z_{10} \\
		y_{12}=\sqrt{h_{u}}x_{1}+z_{12} 
	\end{align}
\end{subequations}
where the noise terms are represented by $z_{10}\sim \mathcal{CN}(0,\sigma^2)$ and $z_{12}\sim \mathcal{CN}(0,\sigma_2^2)$, respectively. Due to \textit{energy causality constraints} \cite{policies}, $U_1$ cannot use the harvested energy during $t_1$ until $t_2$ starts, and hence the energy constraint
\begin{equation}\label{s1_e1}
	P_1^{(1)}t_1 \leq E_1= X_1 t_0
\end{equation}
is imposed. On the other hand, $U_2$ remains inactive during $t_1$ and harvests energy the RF signal transmitted by $U_1$ as well as natural energy. Since $U_1$'s message is useless for $U_2$, all received RF energy is used for EH. %Meanwhile $U_1$ has harvested energy $E_{t1}^{(1)}=X_1t_1$ and $U_2$ has harvested energy $E_{t1}^{(2)}=X_2 t_1+\eta P_1^{(1)} h_{3}t_1$ at the end of $t_1$.

During $t_{2}$, $U_2$ becomes active and broadcasts RF signals to $U_1$ and $D$ with average power $P_2^{(2)}$. Let $x_{2}$ be the complex baseband signal transmitted by $U_2$ and $\mathbb{E}[|x_2|^2]=P_2^{(2)}$. The corresponding received signals at $D$ and $U_1$ can be expressed as
\begin{subequations}
	\begin{align}
y_{20}=\sqrt{h_{2}}x_{2}+z_{20}\\
y_{21}=\sqrt{h_{u}}x_{2}+z_{21}
\end{align}
\end{subequations}
where $z_{21}\sim \mathcal{CN}(0,\sigma_1^2)$ represents the noise at $U_1$. And the energy constraint at $U_2$ is given by
\begin{equation}\label{s1_e2}
P_2^{(2)}t_2 \leq X_2 (t_0+t_1)+\eta P_1^{(1)} h_{u}t_1.
\end{equation}
Meanwhile, $U_1$ splits the received RF signal into two streams with PS ratio $\rho_a$, hence $(1-\rho_a) h_{u} P_2^{(2)}$ for ID and $\eta \rho_a h_{u} P_2^{(2)}$ for EH, where the EH part is saved for use during $t_3$.

During $t_{3}$, $U_1$ uses its remaining energy to relay $U_2$'s message to $D$.
The signal transmitted by $U_1$ is denoted as $x_3$ with $\mathbb{E}[|x_3|^2]=P_3^{(1)}$, which is received by $D$ as $y_{30}$, namely
\begin{equation}
y_{30}=\sqrt{h_{1}}x_{3}+z_{30}.
\end{equation}
The associated energy constraint during $t_3$ is given by
\begin{equation}\label{s1_e3}
	P_3^{(1)}t_3 \leq X_1 (t_0+t_1+t_2)-P_1^{(1)}t_1+ \rho_a \eta P_2^{(2)} h_{u}t_2.
\end{equation}
Note that although $U_2$ harvests energy from natural energy during $t_2,t_3$, the energy is stored for use in the next transmission block, and does not contribute to the current transmission block. It is for this reason this energy is not taken into account in the formulations of energy constraints. 

\begin{table}[t]
	\centering
	\caption{ID and EH activities at receivers in S1-A.}
	\label{P1-CA}
	\begin{tabular}{|c|c|c|c|c|c|}
		\hline
		\multirow{2}{*}{}& $U_1\to D$              & $U_1\to U_2$        & $U_2\to D$      & \multicolumn{2}{c|}{$U_2\to U_1$} \\ \cline{2-6} 
		\multirow{-2}{*}{}          & ID                      & EH                  & ID              & ID                   & EH         \\ \hline
		\multicolumn{1}{|c|}{$t_1$} & $U_1$'s message         & \cmark              & \multicolumn{3}{c|}{ \backslashbox{}{}\cellcolor{gray!25}}       \\ \hline
		\multicolumn{1}{|c|}{$t_2$} & \multicolumn{2}{c|}{ \backslashbox{}{} \cellcolor{gray!25}} &  \multicolumn{2}{c|}{$U_2$'s message }      & \cmark     \\ \hline
		\multicolumn{1}{|c|}{$t_3$} & $U_2$'s message         & \cmark              & \multicolumn{3}{c|}{\backslashbox{}{} \cellcolor{gray!25}}       \\ \hline
	\end{tabular}
\end{table}

Finally, in one block $T$, the amount of data transmission achievable by $U_1$ and $U_2$ can be expressed as
\begin{subequations}
	\label{ed_ca_data1}
	\begin{align}
	B_{1A}&=t_{1}\log(1+\frac{P_1^{(1)}h_{1}}{\sigma^2})=t_{1}\log(1+P_1^{(1)}\gamma_1)\\
		B_{2A}&=\min [B_{2A}^{(2)}+B_{2A}^{(1)},B_{2A}^{(u)}] 
	\end{align}
\end{subequations}
respectively, where $\gamma_1=\frac{h_{1}}{\sigma^2}$ \cite{relaycap}. And the amounts of data transmission achievable from $U_2$ to $D$ ($B_{2A}^{(2)}$), from $U_1$ to $D$ ($B_{2A}^{(1)}$), and from $U_2$ to $U_1$ ($B_{2A}^{(u)}$) are given by
\begin{subequations}
	\label{ed_ca_data2}
	\begin{align}
	%\begin{split}
	B_{2A}^{(2)}&=t_{2} \log(1+\frac{P_2^{(2)} h_{2}}{\sigma^2})=t_{2} \log(1+P_2^{(2)}\gamma_2)\\
	B_{2A}^{(1)}&=t_{3} \log(1+\frac{P_3^{(1)} h_{1}}{\sigma^2})=t_{3} \log(1+P_3^{(1)}\gamma_1)	\\
	B_{2A}^{(u)}&=t_{2} \log(1+\frac{(1-\rho_a) P_2^{(2)}h_{u}}{\sigma_1^2})=t_{2} \log(1+\bar{\rho}_a P_2^{(2)}\gamma_u)
	\end{align}
\end{subequations}\label{B2_a}
respectively, where $\gamma_2=\frac{h_{2}}{\sigma^2},\gamma_u=\frac{h_{u}}{\sigma_1^2}$ and $\bar{\rho}_a=1-\rho_a$. It is worth noting that we focus on the case where the source-to-relay channel is better than the source-to-destination channel \cite{relaycap}, so that $\bar{\rho}_a\gamma_u >\gamma_2$ is also imposed. Otherwise the relay $U_1$ should not play a role in the transmission from source $U_2$ to destination $D$.

\textit{Case B:} $U_2$ transmits first.
\begin{table}[ht]
	\centering
	\caption{ID and EH activities at receivers in S1-B.}
	\label{P1-CB}	
	\begin{tabular}{|c|c|c|c|c|c|}
		\hline
	\multirow{2}{*}{} & $U_1\to D$          & $U_1\to U_2$        & $U_2\to D$      & \multicolumn{2}{c|}{$U_2\to U_1$} \\ \cline{2-6} 
		\multirow{-2}{*}{}          & ID                      & EH                  & ID              & ID                   & EH         \\ \hline
		\multicolumn{1}{|c|}{$t_1$} &  \multicolumn{2}{c|}{ \backslashbox{}{} \cellcolor{gray!25}}
	       & \multicolumn{2}{c|}{ $U_2$'s message} &\cmark  \\ \hline
		\multicolumn{1}{|c|}{$t_2$} & $U_2$'s message         & \cmark        & \multicolumn{3}{c|}{ \backslashbox{}{}\cellcolor{gray!25}}    \\ \hline
		\multicolumn{1}{|c|}{$t_3$} & $U_1$'s message         & \cmark              & \multicolumn{3}{c|}{\backslashbox{}{} \cellcolor{gray!25}}       \\ \hline
	\end{tabular}
\end{table}

In this case, each transmission block $T$ also consists of four non-overlapping time intervals, $t_0,t_1, t_2$ and $t_3$. Unlike Case A, however, during $t_1$, $U_2$ transmits RF signals with average power $P_1^{(2)}$. 
%Since $U_2$ only operates during $t_1$, the harvested energy for $U_2$ does not contribute to the remaining phases of current transmission block. Thus we omit these terms. 
During $t_2$, $U_1$ forwards $U_2$'s message with average power $P_2^{(1)}$. During $t_3$, $U_1$ transmits its own message with average power $P_3^{(1)}$. Table \ref{P1-CB} summarizes the ID and EH activities at receivers during $t_1,t_2,$ and $t_3$.

At the end of the transmission block, the amount of data transmitted by $U_1$ and $U_2$, denoted by $B_{1B}$ and $B_{2B}$ respectively, are given by
\begin{subequations}
	\label{ed_cb_data}
	\begin{align}
		%%B_{1B}&=t_{3}\log(1+\frac{P_3^{(1)}h_{1}}{\sigma^2})\\
		B_{1B}&=t_{3}\log(1+P_3^{(1)}\gamma_1)\\
		B_{2B}&=\min [B_{2B}^{(2)}+B_{2B}^{(1)},B_{2B}^{(u)}] 	\\
	%%	B_{2B}^{(2)}&=t_{1} \log(1+\frac{P_1^{(2)} h_{2}}{\sigma^2})\\
		B_{2B}^{(2)}&=t_{1} \log(1+P_1^{(2)} \gamma_2)\\
		%%B_{2B}^{(1)}&=t_{2} \log(1+\frac{P_2^{(1)} h_{1}}{\sigma^2})\\
		B_{2B}^{(1)}&=t_{2} \log(1+P_2^{(1)}\gamma_1)\\
		%%B_{2B}^{(3)}&=t_{1} \log(1+\frac{(1-\rho_b)P_1^{(2)} h_{3}}{\sigma_1^2})
		B_{2B}^{(u)}&=t_{1} \log(1+\bar{\rho}_b P_1^{(2)} \gamma_u)
	\end{align}
\end{subequations} 
where $B_{2B}^{(2)},B_{2B}^{(1)},B_{2B}^{(u)}$ denote the amounts of data transmission achievable from $U_2$ to $D$, from $U_1$ to $D$, and from $U_2$ to $U_1$, respectively. $\rho_b$ denotes the PS ratio at $U_1$ and $\bar{\rho}_b=1-\rho_b$ with $\bar{\rho}_b\gamma_u >\gamma_2$. The energy constraints imposed on the active users are given by
\begin{subequations}
	\label{ed_cb_p}
	\begin{align}
		P_1^{(2)}t_1& \leq X_2 t_0\\
		P_2^{(1)}t_2& \leq X_1 (t_0+t_1)+\eta \rho_b P_1^{(2)} h_{u}t_1\\
		P_3^{(1)}t_3& \leq X_1 (t_0+t_1+t_2)-P_2^{(1)}t_2+ \rho_b \eta P_1^{(2)} h_{u}t_1
	\end{align}
\end{subequations} 

\begin{figure}[t]
	\centering
	\includegraphics[width = 0.6\textwidth]{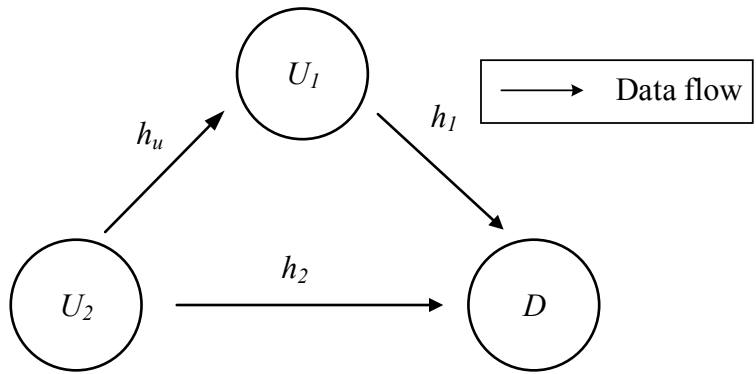}
	\caption{S2: Cooperative transmission model with data cooperation only.}
	\label{fig:dc}
\end{figure}

\textit{ Scenario 2}: Only DC is applied, see Fig. \ref{fig:dc}.

%Cooperative transmission with data cooperation (CTDC): 
In this scenario, the RF signals received at the relay node are utilized for ID without EC, i.e., $\rho_a=\rho_b=0$ and $\eta=0$. Therefore, the data throughputs and energy constraints for Cases A and B can readily be obtained by substituting $\rho_a=\rho_b=0,\eta=0$ into (\ref{s1_e1}), (\ref{s1_e2}) and (\ref{s1_e3}) - (\ref{ed_cb_p}). The ID and EH activities during $t_1,t_2$ and $t_3$ are summarized in Table \ref{P2}. 

\begin{table}[ht]
	\centering
	\caption{ID activities at receivers in S2-A and S2-B with only DC.}
	\label{P2}
	\begin{tabular}{|c|c|c|c|c|}
		\hline
		\multicolumn{2}{|c|}{\multirow{2}{*}{}} & $U_1\to D$      & $U_2\to D$       & $U_2\to U_1$      \\ \cline{3-5} 
		\multicolumn{2}{|c|}{}                  & ID              & ID               & ID       \\ \hline
		\multirow{3}{*}{S2-A}      & $t_1$      & $U_1$'s message & \multicolumn{2}{c|}{\backslashbox{}{} \cellcolor{gray!25}}                \\ \cline{2-5} 
		& $t_2$      &          \backslashbox{}{} \cellcolor{gray!25}       & \multicolumn{2}{c|}{$U_2$'s message} \\ \cline{2-5} 
		& $t_3$      & $U_2$'s message & \multicolumn{2}{c|}{\backslashbox{}{} \cellcolor{gray!25}}         \\ \hline
		\multirow{3}{*}{S2-B}      & $t_1$      &         \backslashbox{}{} \cellcolor{gray!25}       & \multicolumn{2}{c|}{$U_2$'s message} \\ \cline{2-5} 
		& $t_2$      & $U_2$'s message & \multicolumn{2}{c|}{\backslashbox{}{} \cellcolor{gray!25}}                \\ \cline{2-5} 
		& $t_3$      & $U_1$'s message & \multicolumn{2}{c|}{\backslashbox{}{} \cellcolor{gray!25}}                \\ \hline
	\end{tabular}
\end{table}

\begin{figure}[ht]
	\centering
	\includegraphics[width = 0.6\textwidth]{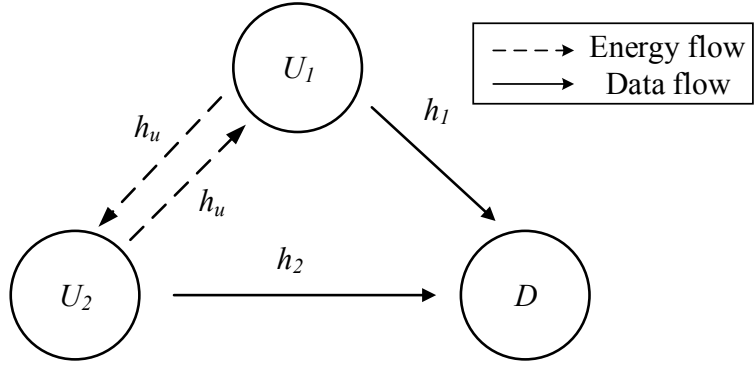}
	\caption{S3: Cooperative transmission model with energy cooperation only.}
	\label{fig:ec}
\end{figure}

\textit{Scenario 3}: Only EC is applied. See Fig. \ref{fig:ec}.

Each transmission block $T$ in this scenario consists of three non-overlapping time intervals $t_0,t_1$, and $t_2$. The two users broadcast their own message during $t_1$ and $t_2$, respectively. However, all RF signals received at the relay node are utilized for EH but with no data relay. The ID and EH activities at receivers are summarized in Table \ref{P3}.

% Please add the following required packages to your document preamble:
% \usepackage{multirow}
\begin{table}[ht]
	\centering
	\caption{ID and EH activities at receivers in S3-A and S3-B with only EC.}
	\label{P3}
	\begin{tabular}{|c|c|c|c|c|c|}
		\hline
		\multicolumn{2}{|c|}{\multirow{2}{*}{}} & $U_1\to D$      & $U_1\to U_2$ & $U_2\to D$      & \multicolumn{1}{c|}{$U_2\to U_1$} \\ \cline{3-6} 
		\multicolumn{2}{|c|}{}                  & ID              & EH           & ID              & \multicolumn{1}{c|}{EH}           \\ \hline
		\multirow{2}{*}{S3-A}      & $t_1$      & $U_1$'s message & \cmark       & \multicolumn{2}{c|}{\backslashbox{}{} \cellcolor{gray!25}}               \\ \cline{2-6} 
		& $t_2$      & \multicolumn{2}{c|}{\backslashbox{}{} \cellcolor{gray!25}}      & $U_2$'s message & \cmark                            \\ \hline
		\multirow{2}{*}{S3-B}      & $t_1$      & \multicolumn{2}{c|}{\backslashbox{}{} \cellcolor{gray!25}}          & $U_2$'s message & \cmark                            \\ \cline{2-6} 
		& $t_2$      & $U_1$'s message & \cmark       & \multicolumn{2}{c|}{\backslashbox{}{} \cellcolor{gray!25}}                               \\ \hline
	\end{tabular}
\end{table}

Under this circumstance, the achievable data transmission and associated energy constraints in Cases A and B can be expressed by
\begin{subequations}
	\label{ec_data_p}
	\begin{align}
	\label{ec_1}
%%	B_{1A}&=t_1\log(1+\frac{P_1^{(1)}h_{1}}{\sigma^2}) \quad P_1^{(1)}t_1\leq X_1t_0 \\
B_{1A}&=t_1\log(1+P_1^{(1)}\gamma_1) \quad P_1^{(1)}t_1\leq X_1t_0 \\
	\label{ec_2}
	%%B_{2A}&=t_2\log(1+\frac{P_2^{(2)}h_{2}}{\sigma^2}) \quad P_2^{(2)}t_2\leq X_2(t_0+t_1)+\eta P_1^{(1)} h_{3}t_1 \\
	B_{2A}&=t_2\log(1+P_2^{(2)}\gamma_2) \quad P_2^{(2)}t_2\leq X_2(t_0+t_1)+\eta P_1^{(1)} h_{u}t_1 \\
	\label{ec_3}
	%%B_{1B}&=t_2\log(1+\frac{P_2^{(1)} h_{1}}{\sigma^2}) \quad P_2^{(1)}t_2\leq X_1(t_0+t_1) + \eta P_1^{(2)} h_{3}t_1 \\
	B_{1B}&=t_2\log(1+P_2^{(1)} \gamma_1) \quad P_2^{(1)}t_2\leq X_1(t_0+t_1) + \eta P_1^{(2)} h_{u}t_1 \\
	\label{ec_4}
	%%B_{2B}&=t_1\log(1+\frac{P_1^{(2)} h_{2}}{\sigma^2}) \quad P_1^{(2)}t_1\leq X_2t_0 
	B_{2B}&=t_1\log(1+P_1^{(2)}\gamma_2) \quad P_1^{(2)}t_1\leq X_2t_0 
	\end{align}
\end{subequations}

\textit{Scenario 4}: Neither DC nor EC occurs.

Here, $U_1$ and $U_2$ transmit their own messages to $D$ directly during $t_1$ and $t_2$. The data throughputs and energy constraints in S4-A and B can readily be established by substituting $\eta=0$ into (\ref{ec_data_p}).

%% file: maxthrpt.tex
\section{Convex Problem Formulation}\label{sec:maxthrpt}
Given natural energy arrival rates at the two user nodes and channel conditions, we seek to determine an optimal scenario in which the distribution of time intervals $\{t_i\}$, power allocations over these time intervals, and data throughputs at the user nodes are jointly optimized so as to maximize the weighted sum-throughput over the present transmission block subject to the constraints imposed for the chosen scenario. We focus on weighted sum-throughput because it helps to give more attention to the sensor node with higher priority in a practical network. An alternative goal instead of weighted sum-throughput maximization is to maximize the minimum throughput of the two users \cite{wpcn}. In this section we present two sets of problem formulations. Based on the analysis and models established in Section \ref{sec:sys_model}, the joint optimization problem is formulated for each EC-DC scenario. As will be shown below, these optimization problems turn out to be convex after a simple variable substitution, hence admit reliable algorithms. The optimal scenario can therefore be identified by evaluating the solutions obtained. Furthermore, based on a second-order approximation of the two-variable \emph{``logarithmic perspective"}, we derive a set of approximate problem formulations. The simplified quadratic formulations remain convex. Both problem formulations, original and approximate, have fast solutions which are developed in Section \ref{sec:algorithm}.
\subsection{Convex formulations based on the models from Section II} \label{sec:3.1}
\emph{(1) Weighted sum-throughput maximization}

The objective here is to maximize the weighted sum throughput over one transmission block with given throughput weights $\bm{w}=[w_1,w_2]$, by jointly optimizing power and time allocations for each of the four transmission scenarios. The sensor node $U_i$ with larger weight $w_i$ has higher priority.

In Scenario S1, the design variables are time allocation $\bm{t}=[t_0,t_1,t_2,t_3]$, power allocations $\bm{P}_A=[P_1^{(1)},P_2^{(2)},P_3^{(1)}],\bm{P}_B=[P_1^{(2)},P_2^{(1)},P_3^{(1)}]$, and throughputs $\bm{B}_A=[B_{1A},B_{2A},B_{2A}^{(1)},B_{2A}^{(2)},B_{2A}^{(3)}]$, $\bm{B}_B=[B_{1B},B_{2B},B_{2B}^{(1)},B_{2B}^{(2)},B_{2B}^{(3)}]$. Based on the model for S1 in Section \ref{sec:sys_model}, the problems at hand for S1-A and B can be formulated, respectively, as
\begin{subequations}
	\label{P1o_A}
	\begin{align}
	\quad \underset{\bm{t},\bm{P}_A,\bm{B}_A}{\text{maximize}} \quad &w_1B_{1A}+w_2 B_{2A}\\
	\text{s.t.} \quad &B_{2A}\leq B_{2A}^{(2)}+B_{2A}^{(1)}\\
	&B_{2A}\leq B_{2A}^{(u)}\\
	& (\ref{s1_e1}),(\ref{s1_e2}),(\ref{s1_e3}),(\ref{ed_ca_data2})\\ \label{t01}
	&t_0+t_{1}+t_{2}+t_3= 1\\
	&t_i\geq 0,i=0,1,2,3
	\end{align}	
\end{subequations}
and
\begin{subequations}
	\label{P1o_B}
	\begin{align}
	\quad \underset{\bm{t},\bm{P}_B,\bm{B}_B}{\text{maximize}} \quad &w_1 B_{1B}+ w_2 B_{2B}\\
	\text{s.t.}  \quad &B_{2B}\leq B_{2B}^{(2)}+B_{2B}^{(1)}\\
	&B_{2B}\leq B_{2B}^{(u)}\\
	 &(\ref{ed_cb_data}),(\ref{ed_cb_p})\\\label{t02}
	&t_0+t_{1}+t_{2}+t_3= 1\\
		&t_i\geq 0,i=0,1,2,3
	\end{align}	
\end{subequations}
%where $B_{1B}=t_{3}\log(1+P_3^{(1)}\gamma_1)$, respectively.

We note that the objective functions in (\ref{P1o_A}) and (\ref{P1o_B}) are in the form $t\log(1+\gamma P)$ which is not concave with respect to $t$ and $P$. However, by using variable substitution $y=Pt$ \cite{usercooperation} and considering $\{t,y\}$ as the design variables instead of $\{t,P\}$, the objective assumes the form $t\log(1+\gamma \frac{y}{t})$ which is the \textit{perspective} of concave function $f(x)=\log(1+\gamma x)$. It then follows from Section 3.2.6 of \cite{optimization} that $g(y,t)=t\log(1+\gamma \frac{y}{t}),t>0$ is also concave with respect to variables $t$ and $y$. The same argument applies to the constraints in problems (\ref{P1o_A}) and (\ref{P1o_B}) that the variable change $y=Pt$ assures that all constraints involved are convex. 

Also note that the constraints in (\ref{t01}) and (\ref{t02}) can be used to eliminate $t_0$ from these problems by substituting $t_0$ by $1-(t_1+t_2+t_3)$ and imposing the non-negativity constraint $1-(t_1+t_{2}+t_3) \geq 0$. In doing so, the problems in (\ref{P1o_A}) and (\ref{P1o_B}) are simplified convex problems as follows, where (P1-A) and (P1-B) denote ``Problem for Scenario 1, Case A'' and ``Problem for Scenario 1, Case B'', respectively:
\begin{subequations}
	\label{P1n_A}
	\begin{align}
	\text{(P1-A)}\quad \underset{\bm{t},\bm{y},\bm{B}_{2A}}{\text{maximize}}\quad &w_1 t_{1}\log(1+\gamma_1\frac{y_1}{t_1}) + w_2 B_{2A}\\
\text{s.t.} \quad &B_{2A}\leq t_{2} \log(1+\gamma_2\frac{y_2}{t_2})+t_{3} \log(1+\gamma_1\frac{y_3}{t_3})\\ \label{P1-A-e1}
	&B_{2A}\leq t_{2} \log(1+\bar{\rho}_a\gamma_u\frac{y_2}{t_2})\\\label{P1-A-e2}
	&y_1 \leq X_1 (1-t_1-t_{2}-t_3)\\\label{P1-A-e3}
	&y_2 \leq X_2 (1-t_{2}-t_3)+\eta h_{u}y_1\\\label{P1-A-e4}
	&y_3 \leq X_1 (1-t_3)-y_1+ \eta \rho_a h_{u}y_2\\\label{P1-A-e5}
	&t_1+t_{2}+t_3 \leq 1\\\label{P1-A-e6}
	&t_i\geq 0,y_i\geq 0,i=1,2,3
	\end{align}	
\end{subequations}
\begin{subequations}
	\label{P1n_B}
	\begin{align}
	\text{(P1-B)}\quad \underset{\bm{t},\bm{y},\bm{B}_{2B}}{\text{maximize}} \quad &w_1 t_{3}\log(1+\gamma_1\frac{y_3}{t_3}) + w_2B_{2B}\\
	\text{s.t.} \quad \quad &B_{2B}\leq t_{1} \log(1+\gamma_2\frac{y_1}{t_1})+t_{2} \log(1+\gamma_1\frac{y_2}{t_2})\\ \label{P1-B-e1}
	&B_{2B}\leq t_{1} \log(1+\bar{\rho}_b\gamma_u\frac{y_1}{t_1})\\\label{P1-B-e2}
	&y_1 \leq X_2 (1-t_1-t_{2}-t_3)\\\label{P1-B-e3}
	&y_2 \leq X_1 (1-t_{2}-t_3)+\eta \rho_b h_{u}y_1\\\label{P1-B-e4}
	&y_3 \leq X_1 (1-t_3)-y_2+ \eta \rho_b h_{u}y_1\\\label{P1-B-e5}
	&t_{1}+t_{2}+t_3\leq 1\\
	&t_i\geq 0,y_i\geq 0,i=1,2,3
	\end{align}	
\end{subequations}

For S2, the formulations of the weighted throughput maximization problems are obtained by substituting $\rho_a=\rho_b=0$ and $\eta=0$ into (\ref{P1n_A}) and (\ref{P1n_B}).

For S3-A, it follows from the model in Section \ref{sec:sys_model} 
(see (\ref{ec_data_p})) and Table \ref{P3} that the throughput maximization problem assumes the form
\begin{subequations}\label{P2n_A}
	\begin{align}
	\text{(P3-A)}\quad \underset{\bm{t},\bm{y}}{\text{maximize}} \quad &w_1 t_1\log(1+\gamma_1\frac{y_1}{t_1})+ w_2t_2\log(1+\gamma_2\frac{y_2}{t_2})\\
	\text{s.t.} \quad \quad &y_1\leq X_1 (1-t_1-t_2)\\
	&y_2\leq X_2(1-t_2)+\eta h_{u}y_1 \\
	&t_{1}+t_{2}\leq 1\\
	&t_i\geq 0,y_i\geq 0,i=1,2
	\end{align}	
\end{subequations}
Similarly, the optimization problem for S3-B can be formulated as
\begin{subequations}\label{P2n_B}
	\begin{align}
		\text{(P3-B)}\quad \underset{\bm{t},\bm{y}}{\text{maximize}} \quad &w_1 t_2\log(1+\gamma_1\frac{y_2}{t_2})+ w_2t_1\log(1+\gamma_2\frac{y_1}{t_1})\\
		\text{s.t.} \quad \quad &y_1\leq X_2 (1-t_1-t_2)\\
		&y_2\leq X_1(1-t_2)+\eta h_{u}y_1 \\
		&t_{1}+t_{2}\leq 1\\
		&t_i\geq 0,y_i\geq 0,i=1,2
	\end{align}	
\end{subequations}

The problem formulations for S4, where no cooperation occurs, can be obtained by simply substituting $\eta=0$ into (\ref{P2n_A}) and (\ref{P2n_B}).

\emph{(2) Maximization of least throughput of the two users}

As an alternative goal to the sum-throughput maximization addressed above, we may consider the problem of maximizing the least throughput of the two users. Problem of this type is known as the common-throughput maximization \cite{wpcn}, and is regarded as useful as it facilitates balancing the resources between the two users. Denote the individual throughputs of users $U_1$ and $U_2$ by $B_1$ and $B_2$, respectively, and let $\bm{B}=[B_1,B_2]$, the common-throughput maximization problem can be formulated as
\begin{subequations}
	\begin{align}
	\text{(P5)}\quad	\underset{\bm{t},\bm{y},\bm{B},\bar{B}}{\text{maximize}}\quad &\bar{B}\\ \label{Bi}
		\text{s.t.} \quad &\bar{B}\leq B_i,i=1,2\\
		&\bm{B} \in \Phi, \bm{t} \in \Psi,\bm{y} \in \Omega
	\end{align}	
\end{subequations}
where $\Phi$, $\Psi$ and $\Omega$ are the feasible sets for throughputs $\bm{B}$, time allocation $\bm{t}$ and auxiliary variable $\bm{y}$, respectively, that can be further specified in accordance with a given scenario. Regardless of the specific scenario involved, $B_i$'s in (\ref{Bi}) always assume the form $t\log(1+\gamma \frac{y}{t})$, hence the constraints in (\ref{Bi}) are convex.
\subsection{Convex quadratic formulations}\label{sec:3.2}

\emph{(1) Quadratic approximation of the perspective function}

\begin{figure}[t]
	\centering
	\includegraphics[width = 0.6\textwidth]{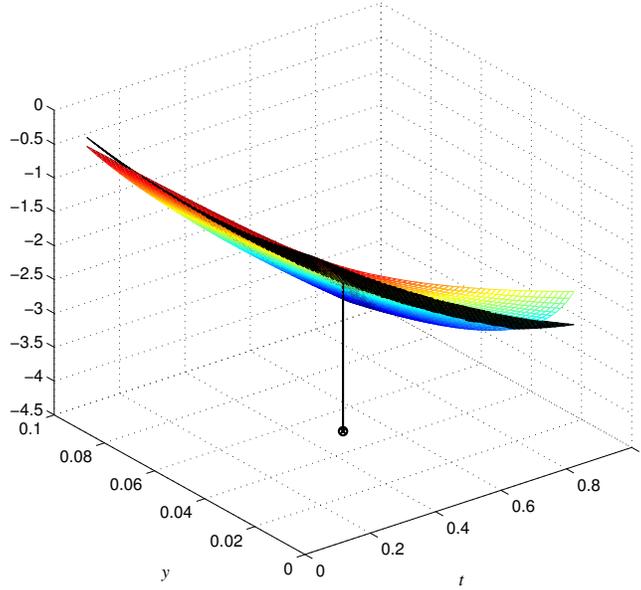}
	\caption{Perspective $l_{\gamma}(t,y)$ (the color surface) versus its quadratic approximation with ($t_k,y_k$) = (0.5, 0.05) (the black surface).}
	\label{fig:quadratic}
\end{figure}

The logarithmic perspective function
\begin{equation}
	l_{\gamma} (t,y) = -t\log(1 + \frac{\gamma y}{t}) 
\end{equation}
encountered in both objective function and constraints is the only non-linear component in the problems formulated in Section 3.1. As such, to a large extent it determines the computational complexity in solving these problems. An effective way to handle $l_{\gamma}(t,y)$ is to build a simple local model of the function at a given iterate ($t_k,y_k$). It turns out that $l_{\gamma}(t,y)$ always has a rank-1 Hessian, hence it admits a very simple convex quadratic model surrounding ($t_k,y_k$) as 
\begin{equation}\label{qua_approx}
	l_{\gamma} (t,y) \approx l_{\gamma} (t_k,y_k) + \bm{g}_k^T\bm{\delta} + \frac{1}{2} (\bm{v}_k^T\bm{\delta})^2 \triangleq l_{\gamma}^{(k)} (t,y)
\end{equation}
where $\bm{g}_k = \begin{bmatrix} -\log(1+\frac{\gamma y_k}{t_k})+\frac{\gamma y_k}{t_k+\gamma y_k} \\ -\frac{\gamma t_k}{t_k+\gamma y_k} \end{bmatrix}$, $\bm{v}_k = \begin{bmatrix} \frac{\gamma y_k}{\sqrt{t_k}(t_k+\gamma y_k)} \\ -\frac{\gamma\sqrt{t_k}}{t_k+\gamma y_k} \end{bmatrix}$ and $\bm{\delta} = [t - t_k, y-y_k]^T$. Fig. \ref{fig:quadratic} depicts the quadratic $l_{\gamma}^{(k)}(t,y)$ (the surface in black) in comparison with the original $l_{\gamma}(t,y)$ (the color surface) over a fairly large region $0.1 \leq t \leq 0.9$ and $0.01 \leq y \leq 0.1$ with $t_k = 0.5$ and $y_k = 0.05$. Over this region, both $l_{\gamma}(t,y)$ and $l_{\gamma}^{(k)}(t,y)$ vary in the range [-4.2475,-0.4615], and the closeness between the two functions in terms of the normalized Frobenius norm of $l_{\gamma}(t,y) - l_{\gamma}^{(k)}(t,y)$ over $251\times 251$ grid points of the region was found to be 0.07.

\emph{(2) Convex quadratic formulations}

Based on (\ref{qua_approx}), convex formulations are simplified counterparts of the problems formulated in Section 3.1 that are obtained by replacing all logarithmic perspective functions involved by quadratic approximation of the form in (\ref{qua_approx}). For illustration, the local quadratic formulation of problem P1-A (see (\ref{P1n_A})) at iterate 
($t_k,y_k$) is given by
\begin{subequations}
	\label{qua_p1}
	\begin{align}
	\text{(P1$^{(k)}$-A)} \quad \underset{\bm{t},\bm{y},\bm{B}}{\text{maximize}}\quad &- w_1 l_{\gamma}^{(k)}(t,y) + w_2 B\\
	\text{s.t.} \quad &B\leq -l_{\gamma_2}^{(k)}(t_2,y_2) t_{2} -l_{\gamma_1}^{(k)}(t_3,y_3)\\ 
	&B\leq -l_{\bar{\rho}_a \gamma_3}^{(k)}(t_2,y_2) \\
	&(\ref{P1-A-e2}),(\ref{P1-A-e3}),(\ref{P1-A-e4}),(\ref{P1-A-e5}), (\ref{P1-A-e6})
	\end{align}	
\end{subequations}
The problem in (\ref{qua_p1}) is a convex quadratically constrained quadratic programming (QCQP) problem. Based on (\ref{P1n_B}) - (\ref{P2n_B}), the quadratic approximation also applies to other scenarios, and QCQP formulations can be constrained for (P1-A), (P1-B), and (P2-B), and QP formulations can be obtained for (P3-A), (P3-B), (P4-A) and (P4-B). It will be shown later in Section \ref{sec:algorithm}, that the algorithm based on approximate QCQP is much faster than that based on the original problems.

%% file: algorithm2.tex
\section{Fast algorithms for optimal energy harvesting}\label{sec:algorithm}
%Energy harvesting (EH) is supposed to be performed in real-time. As such, optimal EH strategies must be constructed fast enough for them to be useful. Reliable and efficient algorithms for constrained convex problems are available, yet the computer codes realizing these algorithms are not designed to take advantages offered by particular problem structures from a particular application. In this section, we propose an enhanced \textit{Newton barrier} algorithm that is tailored to solve the problems formulated in Section \ref{sec:maxthrpt}. Newton methods are well known for fast convergence and high solution accuracy. They are especially suited because of the moderate problem sizes encountered here. A barrier term is incorporated to take care of the constraints and convert the problem at hand into an \emph{unconstrained} convex problem. A line search technique is developed to take full advantage that each Newton direction has to offer so as to accelerate the algorithm considerably. Below the problem P1-A is used as a representative model to illustrate the technical details of the proposed algorithm.

Energy management and data transmission are supposed to be performed in real-time. As such, optimal strategies must be constructed fast enough for them to be useful. Reliable and efficient algorithms for constrained convex problems are available, yet the computer codes realizing these algorithms are not designed to take advantages offered by particular problem structures from a particular application. In this section, we first propose an enhanced \textit{Newton barrier} algorithm that is tailored to solve the original problems formulated in Section \ref{sec:3.1}. Newton methods are well known for fast convergence and high solution accuracy. They are especially suited because of the moderate problem sizes encountered here. A barrier term is incorporated to take care of the constraints and convert the problem at hand into an \emph{unconstrained} convex problem. A line search technique is developed to take full advantage that each Newton direction has to offer so as to accelerate the algorithm considerably. Furthermore, as an alternative approach, we build an iterative algorithm for optimal EH based on the approximate quadratic formulations described in Section \ref{sec:3.2}.
\subsection{The Newton barrier (NB) algorithm for EH}\label{sec:4.1}

\emph{(1) The algorithm}

We consider problem P1-A as a representative formulation to illustrate the technical details of the proposed algorithm. With some notation simplified, P1-A is clearly equivalent to the following convex problem which we shall examine in the rest of Section \ref{sec:4.1}:
\begin{subequations}
	\label{algo_A}
	\begin{align}
	\label{fv}
 \underset{\bm{t},\bm{y},B}{\text{minimize}}\quad &f(\bm{x})= - w_1 t_{1}\log(1+\gamma_1\frac{y_1}{t_1}) - w_2 B\\ \label{c1}
	\text{s.t.} \quad &c_1(\bm{x})=B- t_{2} \log(1+\gamma_2\frac{y_2}{t_2})-t_{3} \log(1+\gamma_1\frac{y_3}{t_3}) \leq 0\\ \label{c2}
	 &c_2(\bm{x})=B-t_{2} \log(1+\beta_3\frac{y_2}{t_2}) \leq 0\\ \label{c3}
	 &c_3(\bm{x})=y_1- X_1 (1-t_1-t_{2}-t_3) \leq 0\\ \label{c4}
	 &c_4(\bm{x})=- \beta_1 y_1 + y_2  -X_2 (1-t_{2}-t_3)\leq 0 \\ \label{c5}
	&c_5(\bm{x})= y_1 - \beta_2 y_2 + y_3 - X_1 (1-t_3) \leq 0\\ \label{c6}
	&c_6(\bm{x})= t_1+t_{2}+t_3 - 1\leq 0 \\  \label{c7}
	&\rho_a < 1- \frac{\gamma_2}{\gamma_u},t_i\geq 0,y_i\geq 0,i=1,2,3
	\end{align}	
\end{subequations}
where $B = B_{2A},\beta_1 = \eta h_u,\beta_2=\eta \rho_a h_{u},\beta_3=\bar{\rho}_a\gamma_3$, and $\bm{x} =[\bm{t},\bm{y},B]$. 

\begin{algorithm}[t]
	{\bf Input}: Energy arrival rates $\{X_i\}$, SNRs $\{\gamma_i\}$, PS ratio $\rho$, weights $\{w_i\}$, $\tau_{\max}$, $\mu$ and tolerance $\epsilon$.\\
	{\bf Output}: Optimal solution $\bm{x}^*$.
	\begin{algorithmic}[1]
		\State Initialize all variables with strictly feasible points $n\gets 0,\tau_n = \tau_0 ,\bm{x}_n^{(0)}\gets\{\bm{B}_0,\bm{t}_0,\bm{y}_0\}$. %\Comment Logarithmic Barrier Method
		\While{$\tau_n<\tau_{\max}$}
		\State \quad Initialize $k\gets 0,\bm{x}_k=\bm{x}_n^{(0)}$  %\Comment Newton's Method
		\While{$||\bm{x}_{k}-\bm{x}_{k-1}||_2>\epsilon$}
		\State Compute gradient $\bm{g}_k$ of $F_{\tau}(\bm{x}_k)$ and Hessian $\bm{H}_k$ of $F_{\tau}(\bm{x}_k)$.
		\State	Compute search direction $\bm{d}_k \gets -\bm{H}_k^{-1}\bm{g}_k$.
		\State Find a step size $\alpha_k$ that minimizes $f(\bm{x}_k+\alpha \bm{d}_k)$ subject to (\ref{c1}) - (\ref{c7}).	
		\State $\bm{x}_{k+1} \gets \bm{x}_k+\alpha_k \bm{d}_k$, $k\gets k+1$.
		\EndWhile
		\State  $\bm{x}_{n+1}^{(0)} \gets \bm{x}_k^*$, $\tau_{n+1}\gets \mu\tau_n $, $n\gets n+1$.
		\EndWhile
		\State $\bm{x}^*\gets \bm{x}_n$
	\end{algorithmic}
	\caption{Newton Barrier Algorithm}
	\label{Newton-Barrier}
\end{algorithm}

The NB algorithm solves (\ref{algo_A}) by iteratively solving the unconstrained problem
\begin{equation}\label{unconstrained}
\text{minimize} \quad F_{\tau}(\bm{x})=-w_1 t_{1}\log(1+\gamma_1\frac{y_1}{t_1})-w_2 B- \frac{1}{\tau}(\sum_{j=1}^{6}\log(-c_j(\bm{x}))+\sum_{i=1}^{3}\log (t_i)+\log(y_i))
\end{equation}
where $\tau>0$ is a barrier parameter that increases as the iteration proceeds so that minimizing $F_{\tau}(\bm{x})$ gradually becomes the same as minimizing the original objective function in (\ref{algo_A}), yet the presence of the logarithmic barrier term in (\ref{unconstrained}) assures that the iterates produced remain strictly inside the feasible region defined by (\ref{c1}) - (\ref{c7}). The algorithm starts with a strictly feasible initial $\bm{x}_0$ and an initial $\tau=\tau_0>0$. Newton algorithm is applied to minimize  $F_{\tau_0}(\bm{x})$ and its solution is in turn used to initiate the next minimization of $F_{\tau_1}(\bm{x})$ where $\tau_1=\mu \tau_0$ with $\mu$ a constant factor, say $\mu=10$, to yield a larger $\tau_1$. The iteration continues until the 2-norm difference between two consecutive iterates is less than a prescribed tolerance $\epsilon$.

For a fixed $\tau$, the Newton step assumes the form 
\begin{subequations}
	\begin{align}
		\bm{x}_{k+1} &= \bm{x}_k+\alpha_k \bm{d}_k \\ \label{dk}
		\bm{d}_k&=-(\nabla^2 F_{\tau}(\bm{x}_k))^{-1} \nabla F_{\tau}(\bm{x}_k)
	\end{align}
\end{subequations}
where $\alpha_k$ minimizes $f(\bm{x}_k+\alpha_k \bm{d}_k)$ subject to constraints (\ref{c1}) - (\ref{c7}).

A technique for fast identification of $\alpha_k$ is described in part (2) of this section. The proposed method is summarized in Algorithm \ref{Newton-Barrier}.

\emph{(2) A line search technique for optimal $\alpha_k^*$}\label{opt_a}

The line search that finds a step size $\alpha_k$ to minimize $f(\bm{x}_k+\alpha\bm{d}_k)$ for given $\bm{x}_k$ and $\bm{d}_k$ consists of three steps. Denote $\bm{x}_k=[\bm{t}_k,\bm{y}_k,B_k]$ that strictly satisfies constraints (\ref{c1}) - (\ref{c7}), and $\bm{d}_k$ obtained from (\ref{dk}) as $\bm{d}_k=[\bm{\delta}_t,\bm{\delta}_y,\delta_B]$ with $\bm{\delta}_t=[\delta_{t_1},\delta_{t_2},\delta_{t_3}]$ and $\bm{\delta}_y=[\delta_{y_1},\delta_{y_2},\delta_{y_3}]$. The first step of the line search determines an interval of $\alpha$, $[0,\alpha_I]$, over which $\bm{x}_k+\alpha\bm{d}_k$ satisfies the linear constraints (\ref{c3}) - (\ref{c7}). The value of such $\alpha_I$ can be found by simply finding the largest $\alpha_j$ such that $c_j(\bm{x}_k+\alpha\bm{d}_k)<0$ over $[0,\alpha_j]$ for $3\leq j\leq 6$; then the largest $\hat{\alpha}$ that over $[0,\hat{\alpha}]$, $\bm{x}_k+\alpha\bm{d}_k$ satisfies (\ref{c7}). Because of the linearity of these constraints, it can be readily verified that 
\begin{subequations}
	\begin{align}
		\alpha_3 &= 
		\begin{cases}
		+\infty \quad \text{if } q_3=-X_1(\delta_{t_1}+\delta_{t_2}+\delta_{t_3})-\delta_{y_1} \geq 0\\
		\frac{c_3(\bm{x}_k)}{q_3}\quad \text{otherwise}
		\end{cases}\\
			\alpha_4 &= 
		\begin{cases}
		+\infty \quad \text{if } q_4=-X_2(\delta_{t_2}+\delta_{t_3})+\beta_1\delta_{y_1}-\delta_{y_2} \geq 0\\
		\frac{c_4(\bm{x}_k)}{q_4}\quad \text{otherwise}
		\end{cases}\\
		\alpha_5 &= 
		\begin{cases}
		+\infty \quad \text{if } q_5=-X_1\delta_{t_3}-\beta_1\delta_{y_1}+\beta_2\delta_{y_2}-\beta_1\delta_{y_3} \geq 0\\
		\frac{c_5(\bm{x}_k)}{q_5}\quad \text{otherwise}
		\end{cases}\\
		\alpha_6 &= 
		\begin{cases}
		+\infty \quad \text{if } q_6=\delta_{t_1}+\delta_{t_2}+\delta_{t_3} \geq 0\\
		\frac{c_6(\bm{x}_k)}{q_6}\quad \text{otherwise}
		\end{cases}\\
		\hat{\alpha} &= \min\{\underset{\delta_{t_i}<0}{-\frac{t_i}{\delta_{t_i}}},\underset{\delta_{y_i}<0}{-\frac{y_i}{\delta_{y_i}}},+\infty\}.
	\end{align}
\end{subequations}
It follows that 
\begin{equation} \label{a1}
	\alpha_I = 0.99\min \{\alpha_3,\alpha_4,\alpha_5,\alpha_6,\hat{\alpha}\}
\end{equation}
where the scaling factor 0.99 assures strict feasibility of $\bm{x}_k+\alpha\bm{d}_k$ on $[0,\alpha_I]$ for (\ref{c3}) - (\ref{c7}).

Next, we identify the largest subinterval $[0,\alpha_{II}]$ of $[0,\alpha_{I}]$, where $\bm{x}_k+\alpha\bm{d}_k$ is strictly feasible for constraints (\ref{c1}) and (\ref{c2}). Note that over $[0,\alpha_I]$ the logarithmic terms in (\ref{c1}) and (\ref{c2}) are well defined because the first six components of $\bm{x}_k+\alpha\bm{d}_k$ (that are the $t$ and $y$ components) remain strictly positive due to (\ref{c7}). Also note that both $c_1(\bm{x}_k+\alpha\bm{d}_k)$ and $c_2(\bm{x}_k+\alpha\bm{d}_k)$ are convex functions of $\alpha$ over $[0,\alpha_I]$. This in conjunction with the fact that both $c_1(\bm{x}_k)$ and $c_2(\bm{x}_k)$ are strictly negative (because $\bm{x}_k$ is a strictly feasible iterate) implies that $c_1(\bm{x}_k+\alpha\bm{d}_k)$ and $c_2(\bm{x}_k+\alpha\bm{d}_k)$ each has at most one zero crossing over $[0,\alpha_I]$ and the zero crossings, denoted by $\alpha^{(1)}$ and $\alpha^{(2)}$ for $c_1(\bm{x}_k)$ and $c_2(\bm{x}_k)$, respectively, can be identified by standard bisection which checks the sign of the function at the mid-point of the interval to decide which part of the interval to keep in order to proceed the bisection procedure. Evidently, $\alpha_{II}=0.99\min\{\alpha^{(1)},\alpha^{(2)}\}$ assures that both $c_1(\bm{x}_k+\alpha\bm{d}_k)$ and $c_2(\bm{x}_k+\alpha\bm{d}_k)$ remain strictly feasible for $\alpha$ over $[0,\alpha_{II}]$. 

The final step of the line search looks for a minimizer of the objective function $f(\bm{x}_k+\alpha\bm{d}_k)$ in (\ref{fv}) as a function of $\alpha$ over $[0,\alpha_{II}]$. Because $f(\bm{x}_k+\alpha\bm{d}_k)$ is convex with respect to $\alpha$, it has a unique minimizer over $[0,\alpha_{II}]$, and the minimizer, denoted by $\alpha_k$, can readily be identified using, for example, the golden-section method that only requires a small number of evaluations of the objective function \cite{goldensection}.
\subsection{An iterative algorithm based on local quadratic formulations}
For illustration purpose we consider the problem in (\ref{qua_p1}), which we call P1$^{(k)}$-A as it is obtained through quadratic approximation of the logarithmic perspective functions involved in P1-A and the formulation is valid in a vicinity of the $k$th iterate ($t_k,y_k$).

With a strictly feasible initial point $\bm{x}_0$, the proposed algorithm calls a convex programming (CP) solver to solve the convex QCQP subproblem P1$^{(0)}$-A as formulated in (\ref{qua_p1}) for a global solution denoted as $\bm{x}_0^*$. This $\bm{x}_0^*$ serves as an initial point for the next iteration in that the quadratic approximation is carried out at $\bm{x}_0^*$ and the updated QCQP subproblem is solved to obtain iterate $\bm{x}_1^*$. The iteration continues until $||\bm{x}_k - \bm{x}_{k-1}||_2$ falls below a prescribed tolerance $\epsilon$. 

The reader is referred to Algorithm \ref{algo_2} for a step-by-step outline of this iterative approach. Our numerical experiments have demonstrated that Algorithm \ref{algo_2} converges within a small number (typically in less than 10) iterations. Therefore, the complexity of Algorithm \ref{algo_2} is largely determined by the complexity of the QCQP subproblem. We have developed an interior-point path-following primal-dual algorithm which is tailored to the structure of problem (\ref{qua_p1}) with a closed-form exact line search step. The customized MATLAB code implementing the above algorithm was evaluated in comparison with a CVX-based counterpart, and the average CPU time required by our code was found in the range 0.0086 to 0.01 normalized time units versus 1 time unit by the CVX-based code. Note that the programs are all run on a MacBook Pro with 2.7 GHz Intel Core i5 processor and 8 GB 1867 MHz DDR3 memory.

We remark that Step 3 of Algorithm \ref{algo_2} where a CP solver is called to solve a QCQP subproblem will have to be modified to solve a QP subproblem when scenarios P3 and P4 are examined. The QP subproblem can be solved by an efficient interior-point algorithm \cite{goldensection}. Comparisons of average CPU time required by several available computer code are illustrated in Table \ref{timecomp} where the average CPU required by MATLAB function $\mathrm{fmincon}$ was normalized to one unit.
\begin{algorithm}[t]
	{\bf Input}: Energy arrival rates $\{X_i\}$, SNRs $\{\gamma_i\}$, PS ratio $\rho$, weights $\{w_i\}$ and tolerance $\epsilon$.\\
	{\bf Output}: Optimal solution $\bm{x}^*$.
	\begin{algorithmic}[1]
		\State Initialize all variables with strictly feasible points $k\gets 0, \bm{x}_0^*\gets\{\bm{B}_0,\bm{t}_0,\bm{y}_0\}, \text{dif}_0 >\epsilon$. 
		\While{$\text{dif}_k > \epsilon$}
		\State Use $\bm{x}_k^*$ to initiate a CP solver for P1$^{(k)}$-A in (\ref{qua_p1}). Denote the solution obtained by $\bm{x}_{k+1}^*$. 
		%\State Call a CP solver and obtain the solution $\bm{x}_k^*$ of P1$^{(k)}$-A in (\ref{qua_p1}).
		%\State	Compute search direction $\bm{d}_k \gets \bm{x}_k^* -  \bm{x}_k$.
		\State  $\text{dif}_k=||\bm{x}_{k+1}^*-\bm{x}_{k}^*||_2$, $k\gets k+1$.
		\EndWhile
		\State $\bm{x}^*\gets \bm{x}_k^*$
	\end{algorithmic}
	\caption{An iterative algorithm based on local quadratic formulations}
	\label{algo_2}
\end{algorithm}
%\begin{table}[h]
%	\centering
%	\caption{Comparisons of Average CPU Time Ratio where Average CPU Time \protect\\ Required by MATLAB Function FMINCON was Normalized to 1 Unit}
%	\label{timecomp}
%	\begin{tabular}{|c|c|c|c|c|c|}
%		\hline
%		\multicolumn{3}{|c|}{\textbf{Our proposed methods}} & \multicolumn{3}{c|}{\textbf{Other Available methods}} \\ \hline
%		\textbf{NB}     & \textbf{QCQP (P1\&P2)}    & \textbf{QC (P3\&P4)}    & \textbf{fmincon}  & \textbf{quadpro}  & \textbf{CVX}  \\ \hline
%		0.5517   &    --   &   --     & 1                 &   --    & --    \\ \hline
%	--       & 0.4917    &   --           & 1           &   --       & 55.25            \\ \hline
%		-- &  --  & 0.2169         & 1     & 3.602           & 60.24            \\ \hline
%	\end{tabular}
%\end{table}
\begin{table}[h]
	\centering
	\caption{Comparisons of Average CPU Time Ratio where Average CPU Time \protect\\ Required by MATLAB Function FMINCON was Normalized to 1 Unit}
	\label{timecomp}
	\begin{tabular}{|c|c|c|c|c|c|c|}
		\hline
		\multirow{3}{*}{} & \multicolumn{3}{c|}{\textbf{Our proposed methods}}                                                                                                                & \multicolumn{3}{c|}{\textbf{Other available methods}}                                                                                                             \\ \cline{2-7} 
		& \textbf{\begin{tabular}[c]{@{}c@{}}original model\end{tabular}} & \multicolumn{2}{c|}{\textbf{\begin{tabular}[c]{@{}c@{}}approximate model\end{tabular}}} & \multicolumn{2}{c|}{\textbf{\begin{tabular}[c]{@{}c@{}}original model\end{tabular}}} & \textbf{\begin{tabular}[c]{@{}c@{}}approximate model\end{tabular}} \\ \cline{2-7} 
		& \textbf{NB}                                                         & \textbf{QCQP}                                 & \textbf{QP}                                 & \textbf{fmincon}                              & \textbf{CVX}                             & \textbf{quadpro}                                                       \\ \hline
		\textbf{P1-A}     & 0.5517                                                              & 0.4917                                        & --                                           & 1                                             & 55.25                                    & --                                                                      \\ \hline
		\textbf{P3-A}     & 0.0294                                                                   & --                                             & 0.2169                                      & 1                                             & 60.24                                    & 3.602                                                                  \\ \hline
	\end{tabular}
\end{table}

%% file: simulation.tex
\section{Numerical Results}\label{sec:simulation}
%\subsection{Parameter Settings}
In the numerical study reported below, the path loss exponent was set to $a=2$ and the average signal power attenuation at a reference of 1 unit of distance was set to $\lambda = 1$, hence channel gain $h_i=d_i^{-2}$ for $i=1,2,u$, where $d_u$ denotes the distance between $U_1$ and $U_2$. All receiver noise power was set to $10^{-4}$, and the energy conversion efficiency $\eta=0.75$ \cite{weiheng}. The weights of two users were set to be equal as $w_1=w_2=1$. To better illustrate the performances of S1, a screening of picking an optimal energy PS ratio $\rho_a$ or $\rho_b$ in S1-A or S1-B that maximizes the weighted sum-throughput or common-throughput among PS ratio candidates from 0 to $\rho_{\max}=1-\frac{\gamma_2}{\gamma_u}$ with an increment of 0.1 was conducted. As PS ratio increases, the EH part gets larger while the ID part gets smaller. Through numerical study, the optimal strategy that employs EC or DC or both is searched for different channel conditions and energy harvesting environments in a simple three-node setup. For illustration purpose, the three nodes are assumed to be positioned on one line.
\subsection{Performances versus energy arrival rates}
\begin{figure}[t]	
	\subfloat[Sum-throughput maximization]{\includegraphics[width=0.51\textwidth]{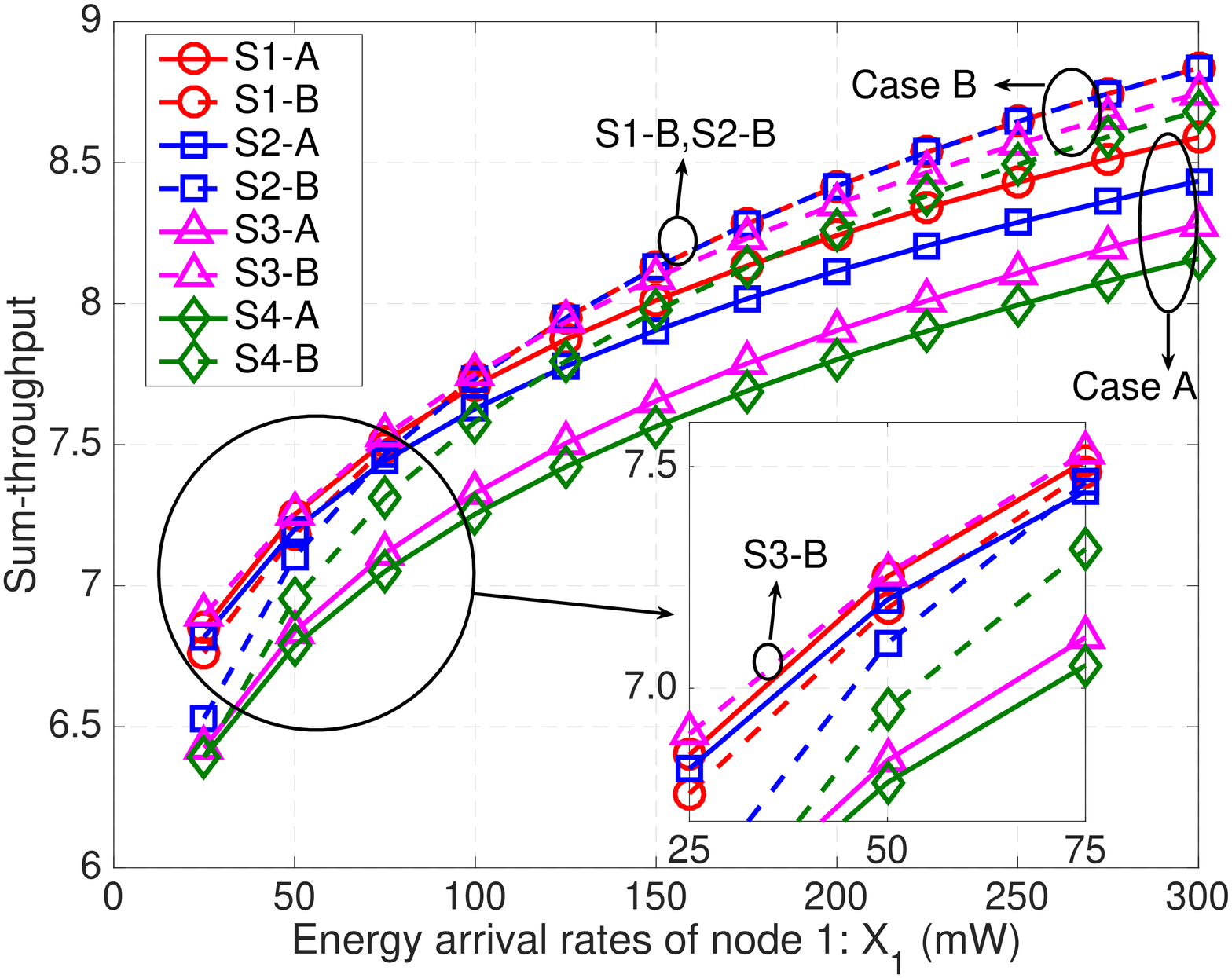}\label{q1c_sum}}
	\subfloat[Common-throughput maximization]{\includegraphics[width=0.51\textwidth]{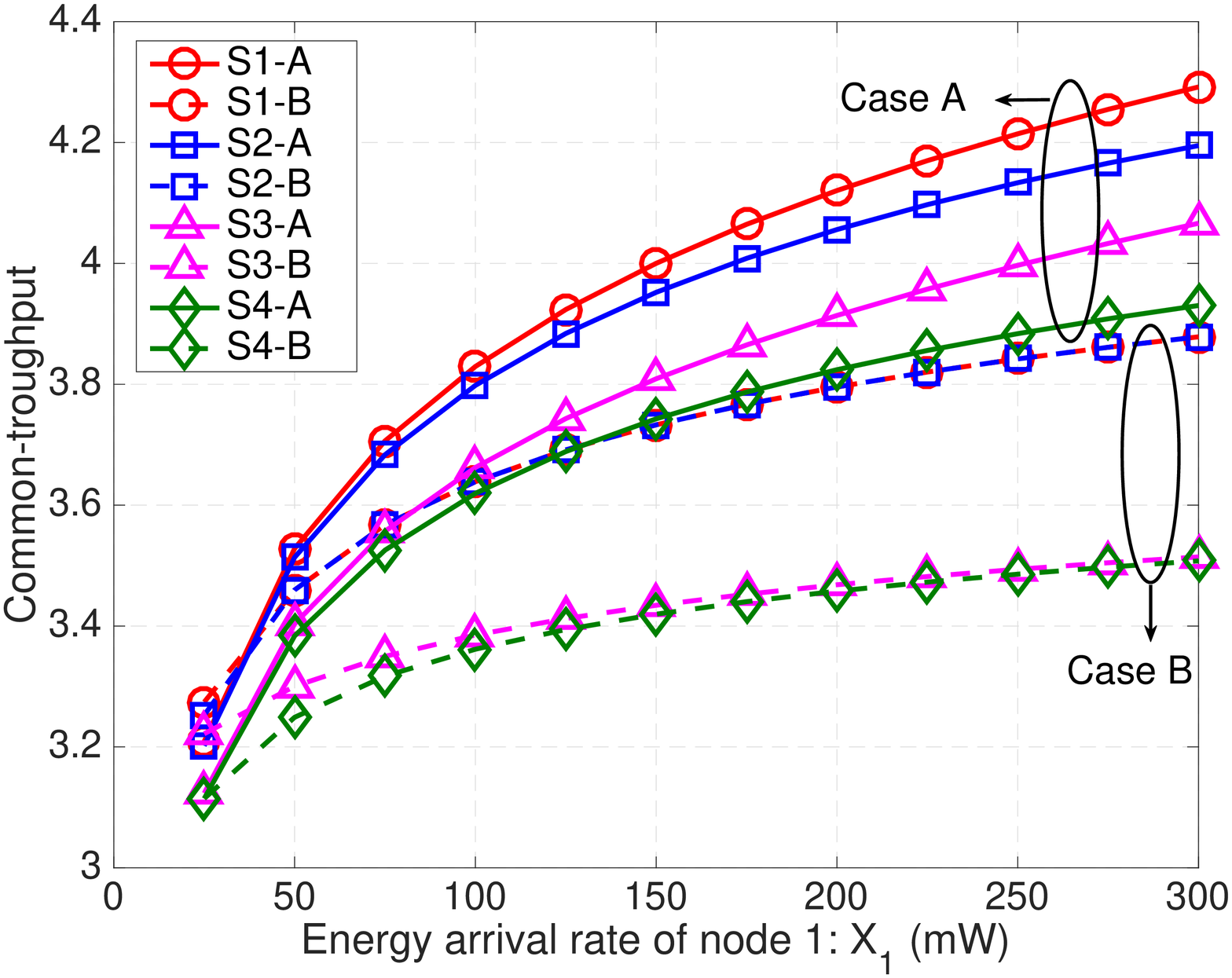}\label{q1c_commnon}}
	\caption{(a) Sum-throughput maximization versus (b) common-throughput maximization in S1 - S4 with $X_2$ fixed to 100 mW.}
	\label{fig:qc_common}
\end{figure}
\begin{figure}[t]
	\centering
	\subfloat[S1-A]{\includegraphics[width=0.51\textwidth]{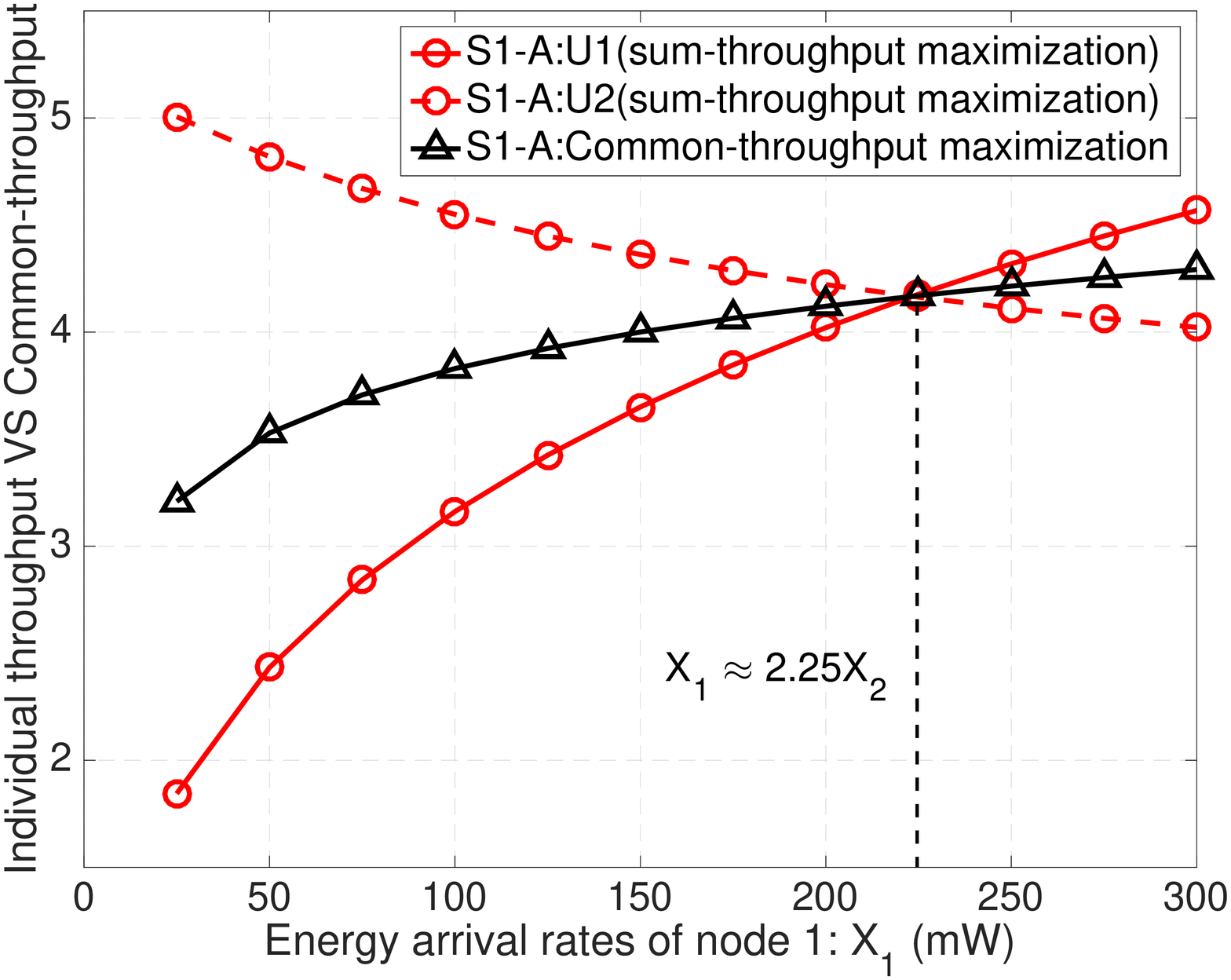}\label{q1c_s1a}}
	\subfloat[S1-B]{\includegraphics[width=0.51\textwidth]{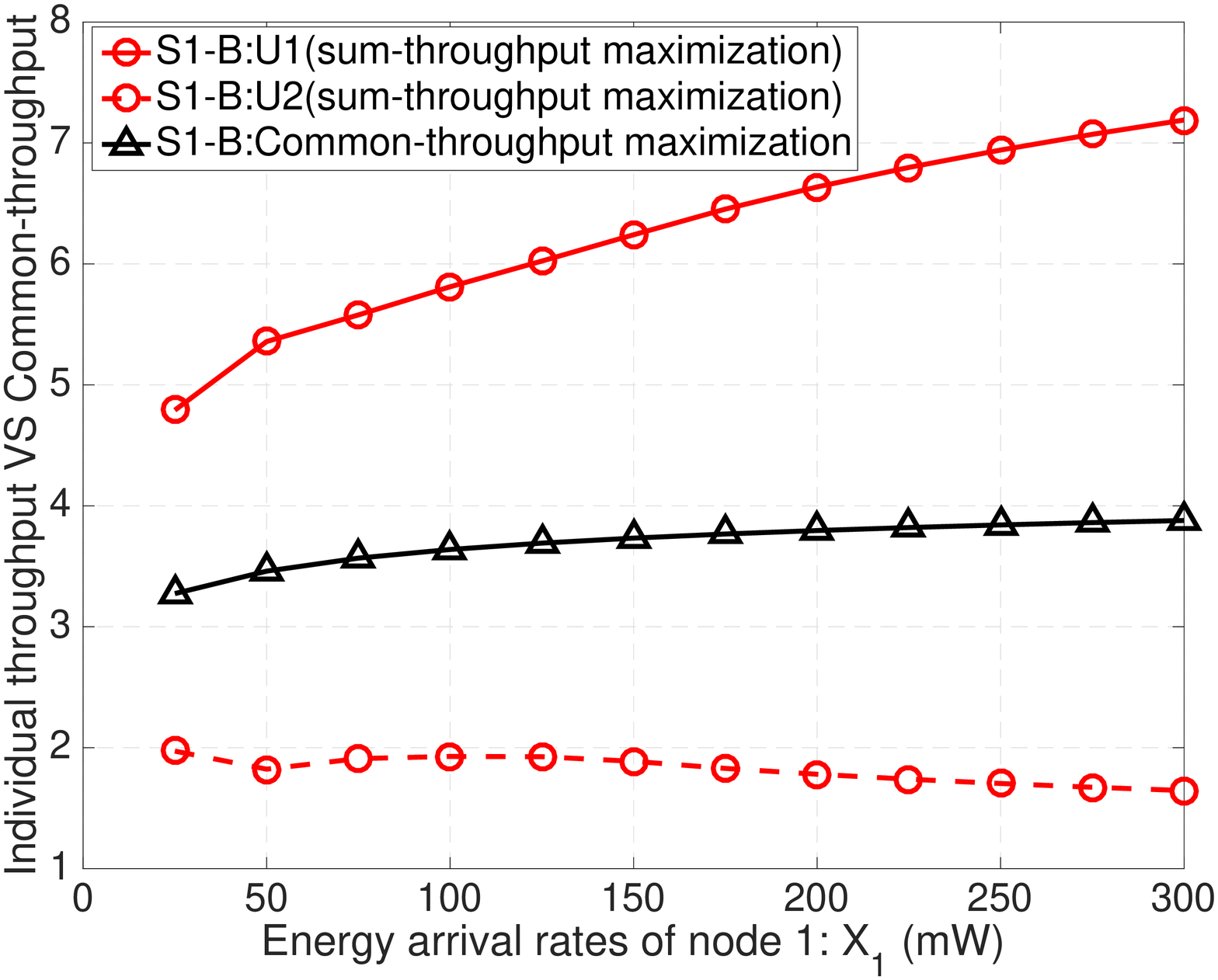}\label{q1c_s1b}}
	\caption{Comparisons of optimal individual throughputs by sum-throughput maximization and the common-throughput by common-throughput maximization in (a) S1-A or (b) S1-B with $w_1=w_2=1$ and $X_2$ fixed to 100 mW.}
	\label{fig:qc_indvd}
\end{figure}
In this part of numerical study, the distance between $U_1$ and $D$ and distance between $U_1$ and $U_2$ were set to $d_1=1$ (unit), and the distance between $U_2$ and $D$ was $d_2=2$. 

By fixing the energy arrival rate $X_2=100$ mW (millijoule per unit of time), the maximum sum-throughputs and common-throughputs in S1 - S4 when $X_1$ increases are compared and shown in Fig. \ref{fig:qc_common}. It is observed in Fig. \ref{fig:qc_common}(a) that when $X_1 > X_2$, i.e., $X_1>100$ mW, the maximum sum-throughputs in Case B ($U_2$ transmits first) dominate those of Case A ($U_1$ transmits first) in all four scenarios. This is because $U_1$ can store more energy for use in the later time intervals. Moreover, S1-B performs very close to S2-B when $X_1 > 150$ mW, which are two best strategies with DC in all scenarios. This means that when the relay node has abundant natural energy to harvest, the harvested energy for $U_1$ is sufficient not only for its own message transmitting, but also for enhancing the throughput of the far user $U_2$. In this case, energy harvesting from $U_2$ by $U_1$ is not needed and Case S1-B degenerates into Case S2-B, which can also be seen from Table \ref{qc_rho} where $\rho_b^*$ is 0 for the given scenario. Also note that when $X_1<0.75X_2$, i.e., $X_1<75$ mW, S3-B is shown to perform slightly better than the others, which indicates that EC should be incurred when the far user $U_2$ has harvested much more energy than $U_1$ and hence to share it with $U_1$. On the other hand, it is illustrated that in Fig. \ref{fig:qc_common}(b) when $X_1 > X_2$, i.e., $X_1>100$ mW, Case A performs better than Case B in all four scenarios for common-throughput maximization, which means $U_1$ should transmit first to balance the performances of two users. In addition, when $X_1>X_2$, i.e., $X_1>100$ mW, S1-A is the best strategy employing energy cooperation and data cooperation, and S2-A is the second best. Both two include DC strategies, which is in agreement with the results in sum-throughput maximization. 

Next, comparisons of optimal individual throughputs achieved by sum-throughput maximization with the common-throughput from common-throughput maximization in S1-A and S1-B are shown in Fig. \ref{fig:qc_indvd}(a) and Fig. \ref{fig:qc_indvd}(b). We see that in S1-A when $X_1<2.25X_2$ shown in Fig. \ref{fig:qc_indvd}(a), the throughput of $U_2$ dominates that of $U_1$. The two throughput trajectories meet at $X_1\approx 2.25X_2$, and then the throughput of $U_1$ exceeds that of $U_2$ when $X_1>2.25X_2$. This indicates that when $U_1$ transmits first and has sufficient natural energy to harvest, $U_1$ contributes more in the sum-throughput maximization. On the other hand, in S1-B shown in Fig. \ref{fig:qc_indvd}(b), the throughput of $U_1$ dominates that of $U_2$ regardless of the value of energy arrival rate. In such case, if $U_2$ transmits first, it cannot take the advantage of EC from $U_1$'s transmission in the current time block, and therefore $U_2$ performs worse than in S1-A. Also observe that the trajectories of common-throughput maximization tend to be flatter hence less sensitive to the changes in energy arrival rates, compared to that of the individual throughputs as $X_1$ or $X_2$ increases, to achieve the balance and equity between two users.

% Please add the following required packages to your document preamble:
% \usepackage{multirow}
\begin{table}[]
	\centering
	\caption{Optimal PS ratios in S1 versus energy arrival rate ratios ${X_1}/{X_2}$}
	\label{qc_rho}
	\begin{tabular}{|c|c|c|c|c|c|c|c|c|c|c|c|c|c|}
		\hline
		\multicolumn{2}{|c|}{$X_1/X_2$}                                                                                                & \textbf{0.25} & \textbf{0.5} & \textbf{0.75} & \textbf{1} & \textbf{1.25} & \textbf{1.5} & \textbf{1.75} & \textbf{2} & \textbf{2.25} & \textbf{2.5} & \textbf{2.75} & \textbf{3} \\ \hline
		\multirow{2}{*}{\textbf{\begin{tabular}[c]{@{}c@{}}sum-throughput\\ maximization\end{tabular}}}    & \textbf{Case A: $\rho_a^*$} & \multicolumn{12}{c|}{0}                                                                                                                                                           \\ \cline{2-14} 
		& \textbf{Case B: $\rho_b^*$} & 0.7           & 0.7          & 0.5           & 0.3        & 0.1            & \multicolumn{7}{c|}{0}                                                                 \\ \hline
		\multirow{2}{*}{\textbf{\begin{tabular}[c]{@{}c@{}}common-throughput\\ maximization\end{tabular}}} & \textbf{Case A: $\rho_a^*$} & 0.1           & \multicolumn{11}{c|}{0}                                                                                                                                           \\ \cline{2-14} 
		& \textbf{Case B: $\rho_b^*$} & 0.4           & \multicolumn{11}{c|}{0}                                                                                                                                           \\ \hline
	\end{tabular}
\end{table}

Numerical values of the optimal PS ratios of Case A and Case B versus energy arrival rate ratios $X_1/X_2$ in S1 are shown in Table \ref{qc_rho}. It follows that PS ratios remain 0 in most cases when natural energy for $U_1$ is sufficient, and the received signals from $U_2$ are all used for ID, which is in agreement with Fig. \ref{fig:qc_common} where DC dominates EC. On the other hand, when $X_1$ is very small ($X_1/X_2 \leq 1.25$), $\rho_b^*>0$ for sum-throughput maximization, while $\rho_a^*$ and $\rho_b^*$ are larger than 0 for common-throughput maximization only when $X_1/X_2 = 0.25$. This demonstrates that when $U_2$ has more natural energy to harvest and $U_2$ transmits first, it takes advantage of EC and shares more energy with $U_1$, especially in sum-throughput maximization.

%which means that decoding more data from $U_2$'s signal is a better choice than harvesting energy for $U_1$. This 

\subsection{Performance versus distance from $U_1$ to $D$}
In this part of numerical study, both energy arrival rates of $U_1$ and $U_2$ were fixed to 100 mW ($X_1=X_2=100$ mW), and $U_1$, $U_2$ and $D$ were located in a line, i.e., $d_u=d_2 - d_1$. In addition, the distance between $U_2$ and $D$ was fixed to $d_2=2$, and $d_1$ varies incrementally from 0.2 to 1.8.

\begin{figure}[t]
	\centering	
	\subfloat[Equal-weighted sum-throughput maximization] {\includegraphics[width=0.51\textwidth]{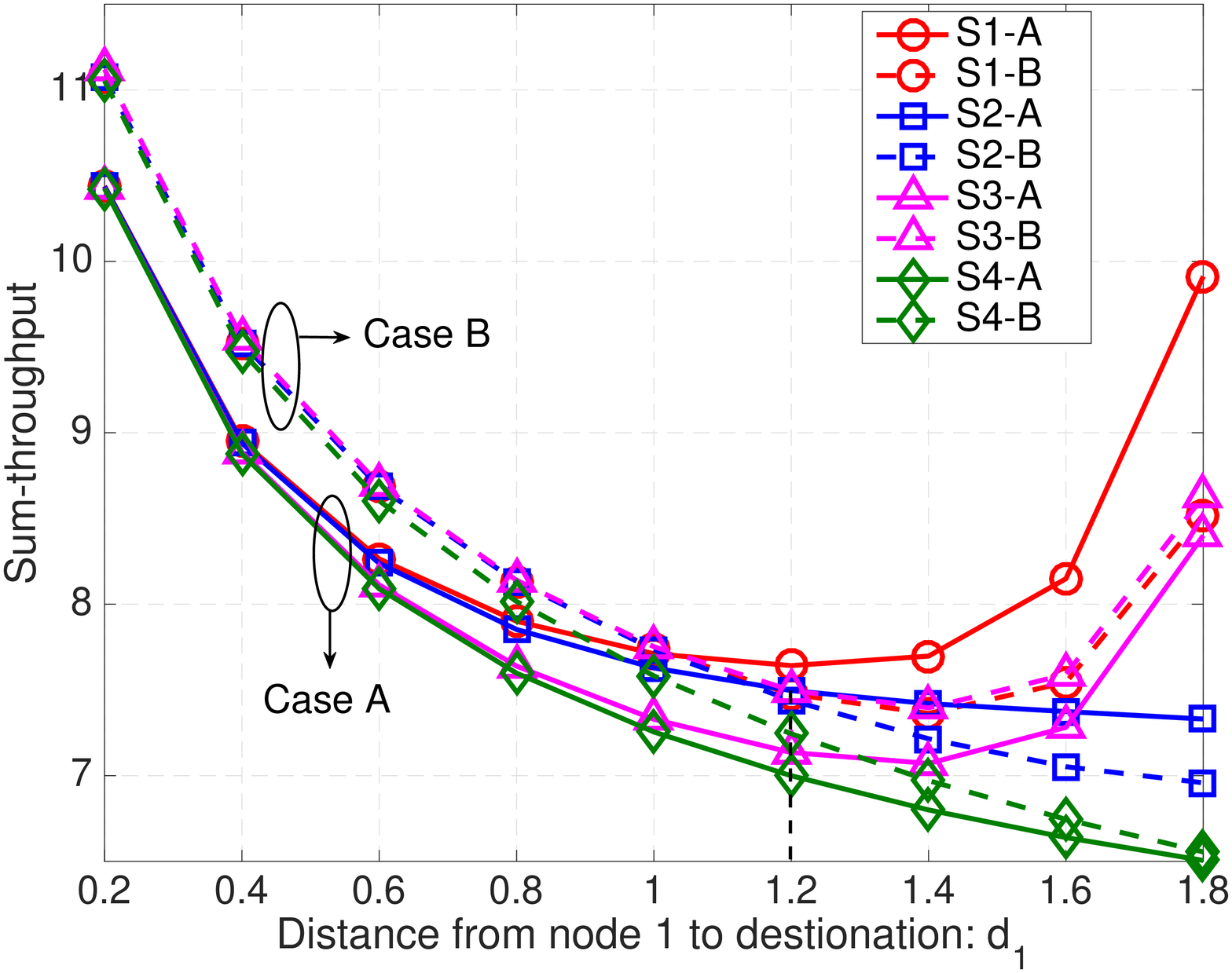}\label{d1c_sum}}
	\subfloat[Common-throughput maximization] {\includegraphics[width=0.51\textwidth]{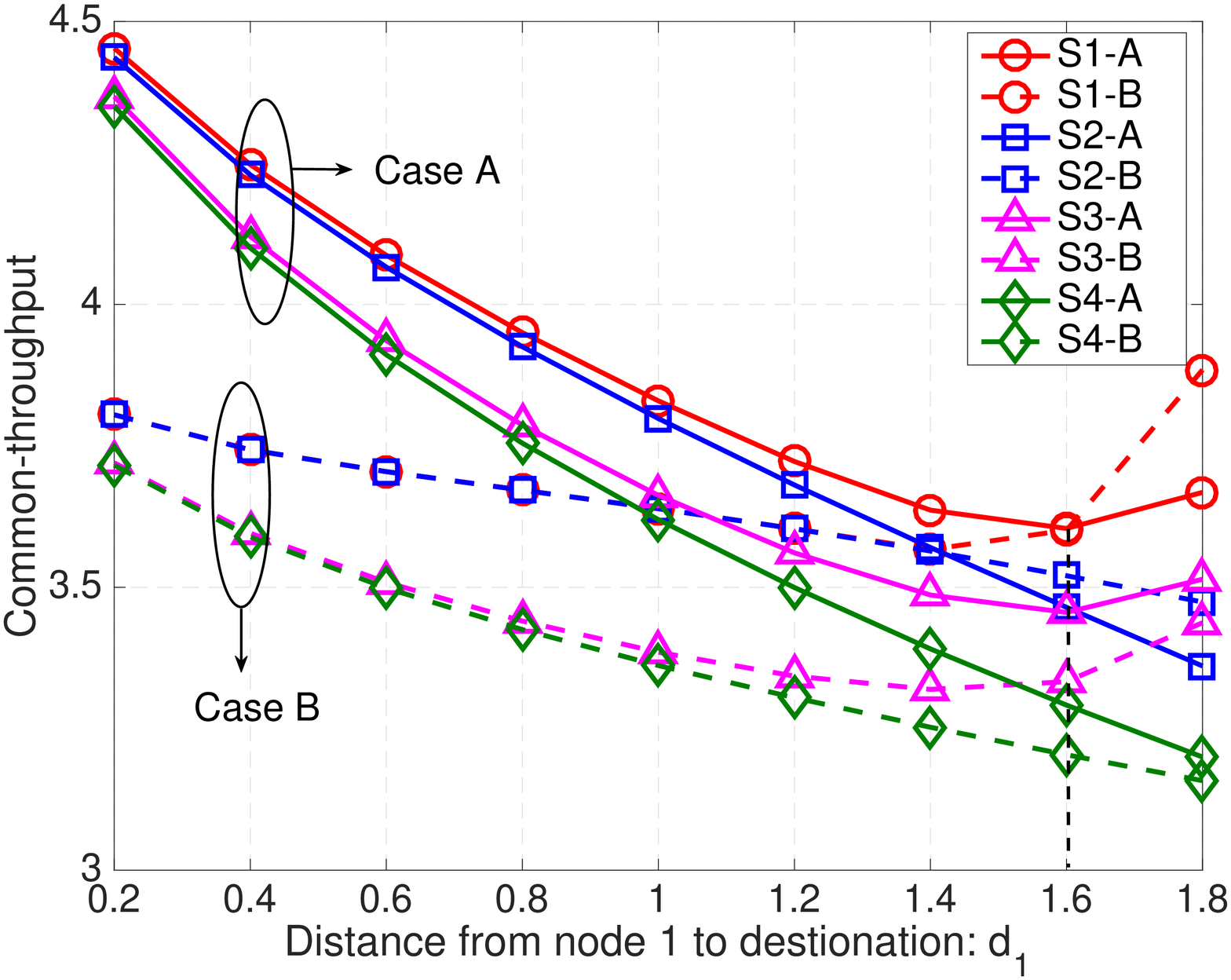}\label{d1c_common}}
	\caption{Comparisons of (a) equal-weighted sum-throughput maximization and (b) common-throughput maximization in S1 - S4 versus distance from $U_1$ to $D$.}
	\label{fig:d1c}
\end{figure}

The results are shown in Fig. \ref{fig:d1c} where optimal sum-throughputs versus distance from $U_1$ to $D$ are compared with the common-throughput obtained by common-throughput maximization. It can be observed that when $U_1$ is far away from $U_2$ ($d_1<1$, $d_u>1$), Case B outperforms Case A in all four scenarios for sum-throughput maximization shown in Fig. \ref{fig:d1c}(a), while Case A outperforms Case B in all four scenarios for common-throughput maximization shown in Fig. \ref{fig:d1c}(b). Also note that in Fig. \ref{fig:d1c}(a) when $U_1$ is far away from $U_2$ ($d_1<0.4$, $d_u>1.6$), four trajectories of sum-throughputs in S1 - S4 for both Case A and Case B are almost coincide with each other. In addition, with $d_1$ increasing from 0.2 to 1.2, both the sum-throughputs and common-throughputs in S1 - S4 are decreasing. However, if two users are close enough, the sum-throughputs in S1 and S3 with EC begin to rise at $d_1=1.2$ ($d_u=0.8$), while common-throughputs in S1 and S3 with EC begin to rise at $d_1=1.6$ ($d_u=0.4$). This is because when two users get closer, the channel condition between the two users improves, which helps to increase the energy harvested from the other user according to (\ref{EH_equ}), i.e., the harvested energy is proportional to channel power gains. This indicates that EC strategies are more adaptive to distance changing relative to DC. On the other hand, in S2 and S4, both sum-throughputs and common-throughputs are found to decrease when two users are getting closer, inferior to S1 and S3. In summary, among the four scenarios, S1 performs the best in both sum-throughput maximization and common-throughput maximization due to the advantages of both EC and DC, while without any cooperation S4 performs the worst.
\begin{table}[ht]
	\centering
	\caption{Optimal PS ratios in S1 versus $d_1$ ($d_2 = 2$)}
	\label{d1c_rho}
	\begin{tabular}{|c|c|c|c|c|c|c|c|c|c|c|}
		\hline
		\multicolumn{2}{|c|}{\textbf{$d_1$}}                                                                                     & \textbf{0.2} & \textbf{0.4} & \textbf{0.6} & \textbf{0.8} & \textbf{1} & \textbf{1.2} & \textbf{1.4} & \textbf{1.6} & \textbf{1.8} \\ \hline
		\multicolumn{2}{|c|}{\textbf{$d_u$}}                                                                                     & \textbf{1.8} & \textbf{1.6} & \textbf{1.4} & \textbf{1.2} & \textbf{1} & \textbf{0.8} & \textbf{0.6} & \textbf{0.4} & \textbf{0.2} \\ \hline
		\multirow{2}{*}{\textbf{\begin{tabular}[c]{@{}c@{}}sum-throughput\\ maximization\end{tabular}}}    & \textbf{Case A: $\rho_a^*$} & \multicolumn{6}{c|}{0}                                                                & 0.1          & 0.4          & 0.5          \\ \cline{2-11} 
		& \textbf{Case B: $\rho_b^*$} & \multicolumn{4}{c|}{0}                              & 0.3        & 0.6          & 0.8          & 0.9          & 0.9          \\ \hline
		\multirow{2}{*}{\textbf{\begin{tabular}[c]{@{}c@{}}common-throughput\\ maximization\end{tabular}}} & \textbf{Case A: $\rho_a^*$} & \multicolumn{6}{c|}{0}                                                                & 0.2          & 0.4          & 0.5          \\ \cline{2-11} 
		& \textbf{Case B: $\rho_b^*$} & \multicolumn{6}{c|}{0}                                                                & 0.2          & 0.6          & 0.7          \\ \hline
	\end{tabular}
\end{table}

Numerical values of the optimal PS ratios in S1 are given in Table \ref{d1c_rho}. It is observed that $\rho_a^*$ for sum-throughput maximization, $\rho_a^*$ and $\rho_b^*$ for common-throughput maximization remain 0 until $d_u$ decreases to 0.6, 0.4 and 0.2. This is because the greater distance between $U_1$ and $U_2$ ($d_u$) leads to poorer channel conditions, hence EC becomes less effective. Similarly, as $d_u$ gets smaller to $d_u=1$, $\rho_b^*$ for sum-throughput maximization begins to rise, and $\rho_b^*$ reaches maximum when $d_u=0.4$ and 0.2. This further shows that if $U_2$ transmits first and channel condition between two users is good, $U_2$ tends to share more energy with $U_1$ to enhance the throughputs.

%\emph{Remark:} 
The above analysis indicates that when $X_1 > X_2$ such that the relay node $U_1$ possesses abundant energy to harvest, DC performs better than EC. We also observe that when $d_1\leq1$ ($d_u \geq 1$), Case A performs better than Case B in common-throughput maximization, while Case B performs better than Case A in sum-throughput maximization. Finally, when two users are close enough ($d_1 > 1$, $d_u<1$), EC strategies are shown to be more optimal than DC.

%% file: Conclusion.tex
\section{Conclusion}\label{sec:conclusion}
In this paper, we have proposed an adaptive cooperative wireless communication network strategy that incorporates energy cooperation (EC) and data cooperation (DC). By examining a three-node network for detailed analysis under the proposal, optimal energy management strategies in terms of maximal weighted sum-throughput or minimum throughput of the two users are investigated in detail for four possible scenarios. Numerical results have shown that whether to include EC and DC depends on the availability of natural energy for harvesting and the topology of the nodes. In many cases, including both EC and DC leads to optimal performance, while EC is more effective when the two users are sufficiently close to each other, and DC should be employed when the near node $U_1$ has abundant natural energy to harvest.

%% file: ms.bbl
% Generated by IEEEtran.bst, version: 1.14 (2015/08/26)
\begin{thebibliography}{10}
\providecommand{\url}[1]{#1}
\csname url@samestyle\endcsname
\providecommand{\newblock}{\relax}
\providecommand{\bibinfo}[2]{#2}
\providecommand{\BIBentrySTDinterwordspacing}{\spaceskip=0pt\relax}
\providecommand{\BIBentryALTinterwordstretchfactor}{4}
\providecommand{\BIBentryALTinterwordspacing}{\spaceskip=\fontdimen2\font plus
\BIBentryALTinterwordstretchfactor\fontdimen3\font minus
  \fontdimen4\font\relax}
\providecommand{\BIBforeignlanguage}[2]{{%
\expandafter\ifx\csname l@#1\endcsname\relax
\typeout{** WARNING: IEEEtran.bst: No hyphenation pattern has been}%
\typeout{** loaded for the language `#1'. Using the pattern for}%
\typeout{** the default language instead.}%
\else
\language=\csname l@#1\endcsname
\fi
#2}}
\providecommand{\BIBdecl}{\relax}
\BIBdecl

\bibitem{industry}
R.~Allan, ``Energy harvesting powers wireless sensor networks in industrial
  apps,'' Available:
  http://www.electronicdesign.com/4g/energy-harvesting-powers-wireless-sensor-networks-industrial-apps,
  Sep. 2012.

\bibitem{EHwsn}
X.~Jiang, J.~Polastre, and D.~Culler, ``Perpetual environmentally powered
  sensor networks,'' in \emph{Proc. 4th International Symposium on Information
  Processing in Sensor Networks (IPSN'05)}, Apr. 2005.

\bibitem{policies}
O.~Ozel, K.~Tutuncuoglu, J.~Yang, S.~Ulukus, and A.~Yener, ``Transmission with
  energy harvesting nodes in fading wireless channels: Optimal policies,''
  \emph{IEEE J. Sel. Areas Commun.}, vol.~29, no.~8, pp. 1732--1743, Sep. 2011.

\bibitem{course516}
C.~K. Ho and R.~Zhang, ``Optimal energy allocation for wireless communications
  with energy harvesting constraints,'' \emph{IEEE Trans. Signal Process.},
  vol.~60, no.~9, pp. 4808--4818, Sep. 2012.

\bibitem{rfsurvey}
X.~Lu, P.~Wang, D.~Niyato, D.~I. Kim, and Z.~Han, ``Wireless networks with
  {R}{F} energy harvesting: A contemporary survey,'' \emph{IEEE Commun. Surveys
  Tuts}, vol.~17, no.~2, pp. 757--789, 2015.

\bibitem{swipt}
R.~Zhang and C.~K. Ho, ``{M}{I}{M}{O} broadcasting for simultaneous wireless
  information and power transfer,'' \emph{IEEE Trans. Wireless Commun.},
  vol.~12, no.~5, pp. 1989--2001, Mar. 2013.

\bibitem{DPS}
L.~Liu, R.~Zhang, and K.-C. Chua, ``Wireless information and power transfer: A
  dynamic power splitting approach,'' \emph{IEEE Trans. Commun.}, vol.~61,
  no.~9, pp. 3990--4001, Sep. 2013.

\bibitem{DPS2}
X.~Zhou, R.~Zhang, and C.~K. Ho, ``Wireless information and power transfer:
  Architecture design and rate-energy tradeoff,'' \emph{IEEE Trans. Commun.},
  vol.~61, no.~11, pp. 4754--4767, Nov. 2013.

\bibitem{wpcn}
H.~Ju and R.~Zhang, ``Throughput maximization in wireless powered communication
  networks,'' \emph{IEEE Trans. Wireless Commun.}, vol.~13, no.~1, pp.
  418--428, Jan. 2014.

\bibitem{wpcnoverview}
S.~Bi, Y.~Zeng, and R.~Zhang, ``Wireless powered communication networks: An
  overview,'' \emph{IEEE Wireless Commun.}, vol.~23, no.~2, pp. 10--18, May
  2016.

\bibitem{wpcnchallenge}
H.~Tabassum, E.~Hossain, A.~Ogundipe, and D.~I. Kim, ``Wireless-powered
  cellular networks: Key challenges and solution techniques,'' \emph{IEEE
  Commun. Mag.}, vol.~53, no.~6, pp. 63--71, Jun. 2015.

\bibitem{energycooperation}
B.~Gurakan, O.~Ozel, J.~Yang, and S.~Ulukus, ``Energy cooperation in energy
  harvesting wireless communications,'' in \emph{Proc. 2012 IEEE Symp. Inf.
  Theory}, Jul. 2012, pp. 965--969.

\bibitem{cooperative}
K.~Ishibashi, H.~Ochiai, and V.~Tarokh, ``Energy harvesting cooperative
  communications,'' in \emph{Proc. IEEE 23rd Int. Symp. Pers., Indoor and
  Mobile Radio Commun. (PIMRC'12)}, 2012.

\bibitem{relay_DC}
C.~Huang, R.~Zhang, and S.~Cui, ``Throughput maximization for the gaussian
  relay channel with energy harvesting constraints,'' \emph{IEEE J. Sel. Areas
  Commun.}, vol.~31, no.~8, pp. 1469--1479, Aug. 2013.

\bibitem{twoway_relay}
K.~Tutuncuoglu, B.~Varan, and A.~Yener, ``Energy harvesting two-way half-duplex
  relay channel with decode-and-forward relaying: Optimum power policies,'' in
  \emph{18th IEEE Int. Conf. Digital Signal Process.}, Jul. 2013.

\bibitem{swiptrelay}
A.~A. Nasir, X.~Zhou, S.~Durrani, and R.~A. Kennedy, ``Relaying protocols for
  wireless energy harvesting and information processing,'' \emph{IEEE Trans.
  Wireless Commun.}, vol.~12, no.~7, pp. 3622--3636, Jul. 2013.

\bibitem{AFDFrelay}
A.~A. Nasir, X.~Zhou, S.~Durrani, and R.~A. Kennedy, ``Wireless-powered relays
  in cooperative communications: Time-switching relaying protocols and
  throughput analysis,'' \emph{IEEE Trans. Commun.}, vol.~63, no.~5, pp.
  1607--1622, May 2015.

\bibitem{DF_multi}
Z.~Ding, S.~M. Perlaza, I.~Esnaola, and H.~V. Poor, ``Power allocation
  strategies in energy harvesting wireless cooperative networks,'' \emph{IEEE
  Trans. Wireless Commun.}, vol.~13, no.~2, pp. 846--860, Feb. 2014.

\bibitem{htc}
H.~Chen, Y.~Li, J.~L. Rebelatto, B.~F. Uch{\^o}a-Filho, and B.~Vucetic,
  ``Harvest-then-cooperate: Wireless-powered cooperative communications,''
  \emph{IEEE Trans. Signal Process.}, vol.~63, no.~7, pp. 1700--1711, Apr.
  2015.

\bibitem{usercooperation}
H.~Ju and R.~Zhang, ``User cooperation in wireless powered communication
  networks,'' in \emph{Proc. IEEE GLOBECOM}, Dec. 2014.

\bibitem{swipt22}
I.~Krikidis, S.~Timotheou, S.~Nikolaou, G.~Zheng, D.~W.~K. Ng, and R.~Schober,
  ``Simultaneous wireless information and power transfer in modern
  communication systems,'' \emph{IEEE Commun. Mag.}, vol.~52, no.~11, pp.
  104--110, Nov. 2014.

\bibitem{relaycap}
A.~Host-Madsen and J.~Zhang, ``Capacity bounds and power allocation for
  wireless relay channels,'' \emph{IEEE Trans. Inf. Theory}, vol.~51, no.~6,
  pp. 2020--2040, Jun. 2005.

\bibitem{optimization}
S.~Boyd and L.~Vandenberghe, \emph{Convex optimization}.\hskip 1em plus 0.5em
  minus 0.4em\relax Cambridge University Press, 2004.

\bibitem{goldensection}
A.~Antoniou and W.-S. Lu, \emph{Practical optimization: Algorithms and
  engineering applications}.\hskip 1em plus 0.5em minus 0.4em\relax Springer,
  2007.

\bibitem{weiheng}
W.~Ni and X.~Dong, ``Energy harvesting wireless communications with energy
  cooperation between transmitter and receiver,'' \emph{IEEE Trans. Commun.},
  vol.~63, no.~4, pp. 1457--1469, Apr. 2015.

\end{thebibliography}
